\documentclass[a4paper,10pt]{amsart}
\usepackage[english]{babel}
\usepackage{eurosym}
\usepackage{amsfonts}
\usepackage{amsmath}
\usepackage{amssymb}
\usepackage{graphicx}
\usepackage{amscd}
\usepackage{color}
\usepackage{float}
\usepackage{tikz}
\usepackage{pstricks}

\usepackage{multirow}
\usepackage{slashbox}
\usepackage{colortbl}
\usepackage{caption}
\usepackage{subcaption}
\newtheorem{theorem}{Theorem}

\newtheorem{lemma}{Lemma}

\numberwithin{equation}{section}

\begin{document}
\title[Mean excess function]{\sc Consistency bands for the mean excess function and application to graphical goodness of fit test for financial data } 
\maketitle
\begin{center}
\author{Gane Samb {\sc Lo}$^{(1,2)}$}, Diadie {\sc Ba}$^{(2)}$, Elhadji {\sc Deme}$^{(2)}$ and Cheikh T.  {\sc Seck}$^{(2,3)}$.\\

\textit{\small $^{(1)}$\textsl{LSA, Universit\'e Pierre et Marie Curie, France.}~~~~~~~~~~~~~~~~~~~~~~~~~~~~~~~~~~~~~~\\
\small $^{(2)}$\textsl{LERSTAD, Universit\'e Gaston Berger, S\'en\'egal.}~~~~~~~~~~~~~~~~~~~~~~~~~~~~~~~~~~~~~~~\\
\small $^{(3)}$\textsl{Universit\'e de Bambey,  S\'en\'egal.}~~~~~~~~~~~~~~~~~~~~~~~~~~~~~~~~~~~~~~~~~~~~~~~~~~~~~~~~~~~~~}
\end{center}
\begin{abstract}
In this paper, we use the modern setting of functional empirical processes and recent techniques on uniform estimation for non parametric objects to derive consistency bands for the mean excess function in the i.i.d. case. We apply our results for modelling financial data, in particular Dow Jones data basis to see how good the Generalized hyperbolic distribution models fit monthly data.\\
\bigskip
\\
\noindent {\bf Keyswords}: \!Mean excess function, Vapnik-Chervonenkis classes, Entropy numbers, Bracketing numbers, Glivenko-Cantelli and Donsker classes, Functional empirical processes, Stochastic processes, Talagrand bounds, Generalized hyperbolic distributions. 
\end{abstract}

\section{Introduction}

Let $X$ be a random variable defined on a probability space $(\Omega,%
\mathcal{A},\mathbb{P}),$ and let $F$ be its distribution function with endpoint 
\begin{equation*}
x_{F}=\sup \{x\in \mathbb{R},F(x)<1\},
\end{equation*}
and let $\bar{F}=1-F$ its survival function.
\\
\bigskip
\\
Throughout the paper we suppose that $\mathbb{E}|X|<\infty $. The mean
excess function $e(u)$ of $X$ is defined by (see, e.g., Kotz and Shanbhag 
\cite{kotz}, Hall and Wellner \cite{HalWel}, Guess and Proschan \cite{guess})

\begin{equation}
e(u)=\mathbb{E}(X-u/X>u)=\left\{ 
\begin{array}{l}
\displaystyle\frac{1}{\bar{F}(u)}\int_{u}^{\infty }\bar{F}(t)dt\ \ \mbox{if}%
\ \ \bar{F}(u)>0\cr\cr\displaystyle0\ \ \ \ \ \ \ \ \ \ \ \ \ \ \ %
\mbox{whenever}\ \ \ \bar{F}(u)=0.\cr%
\end{array}%
\right.  \label{mef}
\end{equation}
\\
\bigskip
\\
A natural way to estimate the mean excess function $e(u)$ is achieved by using the plug-in method, that is replacing the survival function in (\ref{mef}) by its empirical counterpart, as did Yang \cite{yang}.%
\\
\bigskip
\\
Now consider a sequence $X_{1},X_{2},...$ of independent copies of 
$X$. The plug-in estimator of $e(u)$, for $n\geq 1$, is 
\begin{equation}
{e}_{n}(u)=\frac{\sum_{i=1}^{n}(X_{i}-u)\mathbb{I}_{[X_{i}>u]}}{%
\sum_{i=1}^{n}\mathbb{I}_{[X_{i}>u]}}=\frac{\sum_{i=1}^{n}X_{i}\mathbb{I}%
_{[X_{i}>u]}}{\sum_{i=1}^{n}\mathbb{I}_{[X_{i}>u]}}-u, \label{en}
\end{equation}where $\mathbb{I}_{[X>u]}=1$ if $X>u$ and $0$ otherwise.
\\
\bigskip
\\
\newpage For notation convenience, we denote 
$$
\mathbb{P}_{X}(f_{u}) =\int f_{u}(x)dF(x)=\int_{u}xdF(x)$$
and $$\mathbb{P}_{X}(g_{u}) =\int g_{u}(x)dF(x)=\int_{u}dF(x)=\bar{F}(u)
$$
where $f_{u}(x)=x\mathbb{I}_{[x> u]},\ \ g_{u}(x)=\mathbb{I}_{[x >u]},$ and $\mathbb{P}_{X}$ is the probability law of $X$.
\\
\bigskip
\\
 We also denote by $\mathbb{P}_{n}$ the empirical measure associated with the sample $%
X_{1},\cdots ,X_{n}$. We have 
\begin{equation*}
\mathbb{P}_{n}(f_{u})=\frac{1}{n}\sum_{i=1}^{n}X_{i}\mathbb{I}_{[X_{i} > u]}\
\  \text{and}\ \ \ \mathbb{P}_{n}(g_{u})=\frac{1}{n}\sum_{i=1}^{n}\mathbb{I}%
_{[X_{i}> u]}.
\end{equation*}%
Formulae (\ref{mef}) and (\ref{en}) lead to 
\begin{equation*}
e(u)=\left\{ 
\begin{array}{l}
\displaystyle\frac{\mathbb{P}_{X}(f_{u})}{\mathbb{P}_{X}(g_{u})}-u\ \ %
\mbox{if}\ \  u\leq x_{F}\cr\displaystyle0\ \ \ \ \ \ \ \ \ \ \ \ \ \ \ %
\mbox{if}\ \ \ u>x_{F}.\cr%
\end{array}%
\right.
\end{equation*}%
and 
\begin{equation*}
e_{n}(u)=\left\{ 
\begin{array}{l}
\displaystyle\frac{\mathbb{P}_{n}(f_{u})}{\mathbb{P}_{n}(g_{u})}-u\ \ %
\mbox{if}\ \  u\leq X_{n,n}\cr\displaystyle0\ \ \ \ \ \ \ \ \ \ \ \ \ \
\ \mbox{if}\ \ \ u>X_{n,n},  \cr%
\end{array}%
\right
. \end{equation*}
where $\displaystyle X_{n,n}=\max_{1\leq i \leq n}X_i$.\\
\bigskip
\\
One of the most important motivation of the study of the mean excess
function comes from extreme value theory (EVT). Indeed this function $e(u)$ is linear in the threshold $u$ when $F$ is a Generalized Pareto distribution $(\mathcal{G}PD)$ and this is quite a powerful graphical test for such distributions.%
\newline
 By using the Vapnik-Chervonenkis (VC) classes  and the entropy numbers technics, we have been able to establish that the empirical mean excess function $e_n(u)$ converges almost surely and uniformly. We showed that for any $u_1$ less than the upper endpoint of the distribution $F$,
$$  \sup_{u\leq u_{1}} |{e}_{n}(u)-e(u)| \rightarrow 0 \ \ \text{a.s} \ \ \mbox{as} \ \ n\rightarrow \infty.
 $$   Next, by using the modern theory of functional empirical process mainly exposed in \cite{vaart}, we proved that the empirical mean excess function $e_n(u)$ also weakly converges, that is $$\left\{ \sqrt{n}(e_{n}(u)-e(u)),u\in I\right\} \rightarrow^w \left\{ \mathbb{G}(h_{u}),u\in I\right\} . $$
where $\mathbb{G}$ is a Gaussian process and $\{ h_u, u \in I\}$ is a function family to be both precised later.\\
\bigskip
\\
 Furthermore, using Talagrand's inequality (see \cite{talagrand}), and Mason and al. technics (see \cite{EM2010}), we arrived at finding our best achievement: that is finding consistency bands for the mean excess function $e(u)$. Precisely we establish that for any interval $I=[u_0,u_1]$, $u_1$ being less than the upper endpoint of $F$ and for any $\varepsilon>0$, we have for $n$ large 
$$\mathbb{P}\Big(e_{n}(u)-\frac{E_{n}}{\sqrt{n}}<e(u)<e_n(u)+\frac{E_{n}}{\sqrt{n}},u\in I\Big)> 1-\varepsilon, $$
where $(E_n)_{(n\geq 1)}$ is a non-random sequence of real numbers precised in \textbf{Theorem \ref{theo3}} and where $F$ satisfies a very slight condition.\\
\bigskip
\\
These results allowed us to set graphical goodness of fitting test based on the empirical mean excess function and to apply this test to Dow jones data. We found that the Generalized hyperbolic family distribution reveals, itself, to generally fit financial data.\\
In this remainder of the text, we are going to detail this outlined results, to demonstrate them, to make simulations studies about them, and finaly to apply them to financial data.\\ 
\bigskip
\\
The paper is organized as follows. We state uniform almost sure ($a.s$) convergence results in \textsc{Section} \ref{sec2} and finite-distribution and functional normality theorems in \textsc{Section} \ref{sec3}.\\ \textsc{Section} \ref{sec4} is devoted to setting $a.s$ consistency bands for the mean excess function. In \textsc{Section} \ref{sec5}, simulation studies and data driven applications using Dow Jones data are provided. We finish the paper by a concluding section.\\
\bigskip\\
Before we go any further, it is worth mentioning that, in the sequel, all the suprema, taken over $u<u_{1},$ are measurable since the functions of $u$ that
we consider below, are left or right continuous. This means that we are in
the pointwise-measurability scheme. Thus, even when we use the results and
concepts in \cite{vaart}, we do not need exterior either interior
integrals or convergence in outer probability.

\section{Almost Sure Convergence \label{sec2}}
In this section we are going to prove the uniform almost sure convergence of the empirical mean excess function by using Vapnik-Chervonenkis (VC) classes  and bracketing numbers.
\begin{theorem}
\label{theo1} Suppose that $\mathbb{E}\vert g_{u}(X_{1})\vert <\infty$ and $\mathbb{E}\vert f_{u}(X_{1})\vert <\infty, $ 
then 
\begin{equation*}
\sup_{u<x_{F}} |\mathbb{P}_{n}(g_{u})- \mathbb{P}_{X}(g_{u})| \rightarrow 0 \ \ \mbox{a.s} \  \ \, %
\text{as} \ \  n \rightarrow \infty
\end{equation*}%
and 
\begin{equation*}
\sup_{u<x_{F}} |\mathbb{P}_{n}(f_{u})-\mathbb{P}_{X}(f_{u})| \rightarrow 0\ \ \mbox{a.s} \  \ \, %
\text{as} \ \ n \rightarrow \infty.
\end{equation*}%
For any fixed $u_{1}<x_{F},$ for $\mathbb{E}X_{1}^{2}<+\infty$
\begin{equation*}
\sup_{u\leq u_{1}} |{e}_{n}(u)-e(u)| \rightarrow 0 \ \ \text{a.s}\ \ \text{as} \ \ n\rightarrow \infty .
\end{equation*}
\end{theorem}

\bigskip

\noindent \textbf{Proof}. We observe that $\mathcal{F}_{1}=\{g_{u},u<x_{F}\}$
is a class of monotone real functions with values in $[0,1]$. By Theorem 2.7.5 in \cite{vaart}, the
bracketing number $N_{[\,]}(\varepsilon ,\mathcal{F}_{1},L_{r}(Q))$
is finite (bounded by $\exp (K/\varepsilon )$, for every probability measure 
$Q$, any real $r\geq 1$, and a constant $K$ that only depends on $r$).  
Since  
$\mathbb{E}\vert g_{u}(X_{1})\vert <\infty $ for $u<x_{F},$ $\mathcal{F}_{1}$ 
is functional Glivenko-Cantelli class 
in the sense of Theorem 2.4.1 in \cite{vaart}, meaning that 
\begin{equation}
\sup_{u<x_{F}}|\mathbb{P}_{n}(g_{u})-\mathbb{P}_{X}(g_{u})|\rightarrow 0\  \ \mbox{a.s} \  \ \, %
\text{as}\ \  n \rightarrow \infty .\label{eq1}
\end{equation}\\
\bigskip
\\
The class $\mathcal{F}_2=\{f_u, u \in [u_0,u_1] \},$ with $u_1< x_F$, is a
Vapnik-Chervonenkis class with index 
 $V(\mathcal{F}_2)=3$ and its
envelop is $G=\max(\vert f_{u_0}(x)\vert,\vert f_{u_1}(x)\vert$). Then it
satisfies the uniform entropy condition 2.4.1 in \cite{vaart}. Then 
$\mathcal{F}_2$ is a Donsker class and hence it is a Glivenko Cantelli class, that is 

\begin{equation}
\sup_{u<x_{F}}|\mathbb{P}_{n}(f_{u})-\mathbb{P}_{X}(f_{u})|\rightarrow 0 \ \ 
\text{a.s }\ \ \text{ as }\ \ n\rightarrow \infty .\label{eq2}
\end{equation}%
To finish, fix $u_{1}<x_{F}.$ Then for $u\leq u_{1}$ and $n$ large enough,
we have 
\begin{eqnarray*}
{e}_{n}(u)-e(u) &=&\frac{\mathbb{P}_{n}(f_{u})}{\mathbb{P}_{n}(g_{u})}-\frac{%
\mathbb{P}_{X}(f_{u})}{\mathbb{P}_{X}(g_{u})}  \label{decomp} \\
&=&\frac{\mathbb{P}_{n}(f_{u})}{\mathbb{P}_{n}(g_{u})}-\frac{\mathbb{P}%
_{X}(f_{u})}{\mathbb{P}_{n}(g_{u})}+\frac{\mathbb{P}_{X}(f_{u})}{\mathbb{P}%
_{n}(g_{u})}-\frac{\mathbb{P}_{X}(f_{u})}{\mathbb{P}_{X}(g_{u})}  \notag \\
&=&(\mathbb{P}_{n}(g_{u}))^{-1}(\mathbb{P}_{n}(f_{u})-\mathbb{P}_{X}(f_{u}))-%
\mathbb{P}_{X}(f_{u})\times \frac{\mathbb{P}_{n}(g_{u})-\mathbb{P}_{X}(g_{u})%
}{\mathbb{P}_{n}(g_{u})\mathbb{P}_{X}(g_{u})}.  \notag
\end{eqnarray*}%
Then \begin{equation}\label{ingen}
\vert {e}_{n}(u)-e(u)\vert \leq \vert \mathbb{P}_{n}(g_{u})\vert^{-1}\times \vert\mathbb{P}_{n}(f_{u})-\mathbb{P}_{X}(f_{u})\vert+ \vert\mathbb{P}_{X}(f_{u})\vert\times  \frac{\vert\mathbb{P}_{n}(g_{u})-\mathbb{P}_{X}(g_{u})\vert}{\vert \mathbb{P}_{n}(g_{u})\mathbb{P}_{X}(g_{u})\vert}.
\end{equation}  
Let 
\begin{eqnarray}\label{del}
\displaystyle \epsilon _{n} =\sup_{u<x_F}|\mathbb{P}_{n}(f_{u})-\mathbb{P}%
_{X}(f_{u})| \ \ \mbox{and} \ \ \delta _{n} =\sup_{u<x_F}|\mathbb{%
P}_{n}(g_{u})-\mathbb{P}_{X}(g_{u})|.
\end{eqnarray}%
From \eqref{eq1} and \eqref{eq2} above, we have 
\begin{equation*}
\epsilon_n \rightarrow 0 \ \ \text{a.s} \ \ \mbox{and } \ \ \delta_n
\rightarrow 0 \ \ \text{a.s },\ \ \mbox{as}\ \ n \rightarrow \infty.
\end{equation*}

\bigskip

Now for $u\leq u_{1}$ , we have $\mathbb{P}_{X}(g_{u})\geq \mathbb{P}%
_{X}(g_{u_{1}})$ and from \eqref{del} ,%
\begin{eqnarray*}
&&-\delta _{n}\leq \mathbb{P}_{n}(g_{u})-\mathbb{P}_{X}(g_{u})\leq \delta
_{n} \\
&&-\delta _{n}+\mathbb{P}_{X}(g_{u})\leq \mathbb{P}_{n}(g_{u})\leq \delta
_{n}+P_{X}(g_{u}),
\end{eqnarray*}%
since $\mathbb{P}_{n}(g_{u})\geq \mathbb{P}_{X}(g_{u})-\delta_{n}>0$ for $n$ large enough, then $(%
\mathbb{P}_{n}(g_{u}))^{-1}\leq (\mathbb{P}_{X}(g_{u_{1}})-\delta _{n})^{-1}$%
. \\We also have \begin{equation}\label{fuesp}
\vert \mathbb{P}_{X}(f_{u})\vert =\left \vert \int_u x \,dF(x)\right \vert \leq  \left \vert \int_\mathbb{R}x \,dF(x)\right \vert \leq   \int_\mathbb{R}\vert x  \vert dF(x)=\mathbb{E}\vert X \vert=\alpha <\infty.  
\end{equation}
 Thus 
\begin{equation*}
\sup_{u\leq u_{1}}|{e}_{n}(u)-e(u)|\leq \epsilon _{n}\Big[\mathbb{P}%
_{X}(g_{u_{1}})-\delta _{n}\Big]^{-1}+\alpha \Big[\mathbb{P}_{X}(g_{u_{1}})(%
\mathbb{P}_{X}(g_{u_{1}})-\delta _{n})\Big]^{-1}\delta _{n}
\end{equation*}%
and then 
\begin{equation*}
\sup_{u\leq u_{1}} \vert e_{n}(u)-e(u) \vert \rightarrow 0 \ \ \text{a.s} \ \ \mbox{as} \
\ n\rightarrow \infty.\ \ \ \ \ \ \ \ \ \ \ \ \ \ \ \ \ \ \ \ \ \ \ \ \ \ \ \ \ \ \ \ \ \ \ \ \square
\end{equation*}

\section{Asymptotic normality of $\boldsymbol{e_{n}(u)}$\label{sec3}}

In this section, we are concerned with weak laws of the empirical mean excess
process as a stochastic process. Hereafter \ $\left\{ \mathbb{G}(g),g\in 
\mathcal{G}\right\} $ denotes a Gaussian centered functional stochastic
process with variance-covariance function%
\begin{equation*}
\Gamma (g_{1},g_{2})=\int (g_{1}(x)-\mathbb{E}g_{1}(X_{1}))(g_{2}(x)-\mathbb{%
E}g_{2}(X_{1}))dF(x).
\end{equation*}

\begin{theorem}
\label{theo2} Let $X_{1},$ $X_{2},\cdots $ be  iid $rv$'s with common
finite second moment. \\Put $I=[u_{0},u_{1}],$ with $
u_{0}<u_{1}<x_{F}$ and define the functions of $t \in \mathbb{R},$ 
$$h_{u}(t)=\mathbb{P}_{X}(g_{u})^{-1}f_{u}(t)-\mathbb{P}_{X}(f_{u})\mathbb{P}
_{X}^{-2}(g_{u})g_{u}(t) \ \ \mbox{for} \ \ u\in I.$$
Suppose that $F$ is continuous and satisfies%
$$\limsup_{\delta \rightarrow 0} \sup_{\left( v,v-\delta \right) \in I^{2}} 
\left( \frac{ F(v)-F(v-\delta )}{\sqrt{\delta}}\right)^2 =0.$$

\noindent Then the functional empirical processes $\left\{ \mathbb{G}_{n}(g_{u}),u\in I\right\} $ \ and 
$\left\{ \mathbb{G}_{n}(f_{u}),u\in I\right\} $ weakly converge respectively to $\left\{ 
\mathbb{G}(g_{u}),u\in I\right\} $ and $\left\{ \mathbb{G}(f_{u}),u\in
I\right\} $ in $\ell ^{\infty }(I).$\newline

\noindent And $\left\{ \sqrt{n}(e_{n}(u)-e(u)),u\in I\right\} $ weakly
converges to $\left\{ \mathbb{G}(h_{u}),u\in I\right\} .$
\end{theorem}
\bigskip
Before we give the proof,
we need this lemma.
\\
\bigskip
\\
\begin{lemma}
\label{lemg} \bigskip Let $g$ be a finite measurable function defined on $%
\mathbb{R}$ such that $\mathbb{E}g(X_{1})^{2}<\infty $ . Let $u_{0}<u_{1}<x_{F}.$ Define for any fixed $v\in \mathbb{R}$ and $\delta >0$%
\begin{equation*}
\sigma ^{2}(v,\delta )=\int_{v-\delta }^{v}\left( g(x)-\mathbb{E}%
(g(x)\right) ^{2}dF(x).
\end{equation*}%
Let for a fixed $n\geq 1,$ $u\in \mathbb{R}$%
\begin{equation*}
S_{n}(u)=\frac{1}{\sqrt{n}}\sum_{j=1}^{n}\left[ g(X_{j})  \mathbb{I}_{(X_{j}> u)}-%
\mathbb{E}g(X_{j})\mathbb{I}_{(X_{j}> u)}\right] .
 \end{equation*}%
 
\begin{equation*}
 \mbox{If} \ \ \ \sup_{u_0\leq v\leq u_{1}}\frac{\sigma ^{4}(v,\delta )}{\delta}  \rightarrow 0\text{ }as\text{
} \delta \rightarrow 0 \ \ \ \ \mbox{and} \ \ \ \ \sup_{u_{0}\leq x\leq u_{1}}\left\vert g(x)-\mathbb{E}g(X)\right\vert
<\infty ,
\end{equation*}%
then 
\begin{equation*}
\lim_{\delta \rightarrow 0}\sup_{u_{0}\leq v\leq u_{1}}\sup_{n\geq 1}\frac{1%
}{_{\delta }}P(\sup_{v-\delta \leq u\leq v}\left\vert
S_{n}(u)-S_{n}(v)\right\vert \geq \eta )=0.
\end{equation*}
\end{lemma}

\bigskip

\noindent \textbf{Proof of Lemma \ref{lemg}}. We fix $v\in \mathbb{R}$ and
consider $\displaystyle \alpha =\sup_{v-\delta <u<v}\left\vert
S_{n}(u)-S_{n}(v)\right\vert .$ Observe that for $u<v,$%
\begin{equation*}
S_{n}(u)-S_{n}(v)=\frac{1}{\sqrt{n}}\sum_{j=1}^{n}\left\{ g(X_{j})\mathbb{I}_{]u,v]}(X_j)-%
\mathbb{E} g(X_{j})\mathbb{I}_{]u,v]}(X_j) \right\}.
\end{equation*}

\noindent Since for all 
 $\left(
u,v\right) \in \mathbb{R}^{2}$, we have 
\begin{equation*}
\left\vert S_{n}(v)-S_{n}(u)\right\vert \leq \frac{1}{\sqrt{n}}\sum_{j=1}^{n}%
\left[ \left\vert g(X_{j})\right\vert + \left\vert \mathbb{E} g(X_{j})\right\vert %
\right] <\infty,
\end{equation*}

\noindent it comes that $\alpha $ is finite. So for any $\varepsilon >0,$%
\ we can find $\overset{\_}{u}\in \lbrack v-\delta ,v[$ such that,

\begin{equation}\label{ub}
\left\vert S_{n}(\overline{u})-S_{n}(v)\right\vert \geq \alpha -\varepsilon .
\end{equation}

\noindent Now, let $\delta >0$. Define for any $p\geq 1,$ and consider $%
u_{j}(p)=u_j=v-\delta +j\delta/p$, $j=0,...,p.$ \\ Let us prove that for $%
\varepsilon >0,$%
\begin{equation*}
\lim_{p\rightarrow \infty }\max_{0\leq j\leq p}\left\vert
S_{n}(u_{j})-S_{n}(v)\right\vert \geq \alpha -\varepsilon .
\end{equation*}

\noindent For each $p\geq 1,$ let $j$ such that%
\begin{equation*}
u_{j-1}(p)\leq \overset{\_}{u} \leq u_{j}(p).
\end{equation*}

\noindent We have, 
\begin{eqnarray*}
\left\vert S_{n}(u_{j})-S_{n}(v)\right\vert &\geq & \left\vert S_{n}(\overset{\_%
}{u})-S_{n}(v)\right\vert -\left\vert \frac{1}{\sqrt{n}}%
\sum_{i=1}^{n}(g(X_{i})\mathbb{I}_{]\bar{u},u_j(p)]}(X_i)-\mathbb{E}g(X_{i})\mathbb{I}_{]\bar{u},u_j(p)]}(X_i)\right\vert \\
&\geq & \left\vert S_{n}(\overset{\_%
}{u})-S_{n}(v)\right\vert - R_j(p),
\end{eqnarray*}

\noindent by denoting%
\begin{equation*}
R_{j}(p)=\left\vert \frac{1}{\sqrt{n}}%
\sum_{i=1}^{n}(g(X_{i})\mathbb{I}_{]\bar{u},u_j(p)]}(X_i)-\mathbb{E}g(X_{i})\mathbb{I}_{]\bar{u},u_j(p)]}(X_i)\right\vert.
\end{equation*}

\noindent We get from \eqref{ub}
\begin{equation*}
\max_{0\leq j\leq p}\left\vert S_{n}(u_{j})-S_{n}(v)\right\vert \geq \alpha
-\varepsilon +R_{j}(p).
\end{equation*}

\noindent For a fixed $n\geq 1$, $R_{j}(p)$ $%
\rightarrow 0$ as $p\rightarrow \infty $, since the sequence of intervals $(]\bar{u},u_j(p)])_{p\geq 1}$ decreases to the empty set as $p\rightarrow \infty $.
\\
\bigskip
\\

\noindent Next, consider the collection points $\left\{ u_{j}(\ell
),0\leq j\leq p,1\leq \ell \leq p \right\} $ and denote the set of its
distinct values between them as $\left\{ \overset{\_}{u}_{j},1\leq j\leq
m(p)\right\} .$ We still have $\left\vert \overset{\_}{u_{j}}-\overset{\_}{u}%
_{j-1}\right\vert \leq \delta /p.$ And we surely have for any $\varepsilon
>0 $ 
\begin{equation*}
\lim_{p\rightarrow \infty }\max_{0\leq j\leq m(p)}\left\vert S_{n}(\overset{%
\_}{u}_{j})-S_{n}(v)\right\vert \geq \alpha -\varepsilon
\end{equation*}

\noindent and then 
\begin{equation*}
\lim_{p\rightarrow \infty }\max_{0\leq j\leq m(p)}\left\vert S_{n}(\overset{%
\_}{u}_{j})-S_{n}(v)\right\vert \geq \alpha
\end{equation*}

\noindent and finally 
\begin{equation*}
\sup_{p\geq 1}\max_{0\leq j\leq m(p)}\left\vert S_{n}(\overset{\_}{u}%
_{j})-S_{n}(v)\right\vert =\alpha.
\end{equation*}

\noindent By construction, $\displaystyle \max_{0\leq j\leq m(p)}\left\vert S_{n}(\overset{%
\_}{u}_{j})-S_{n}(v)\right\vert $ is non decreasing in $p$. So, by the
Monotone Convergence Theorem, for any fixed $v >0,$ for any $\eta >0,$%
\begin{equation}
\mathbb{P}(\sup_{v-\delta \leq u\leq v}\left\vert S_{n}(u)-S_{n}(v)\right\vert \geq
\eta )=\lim_{p\uparrow \infty }\mathbb{P}(\max_{1\leq j\leq m(p)}\left\vert S_{n}(%
\overset{\_}{u}_{j})-S_{n}(v)\right\vert \geq \eta ).  \label{approx01}
\end{equation}
\noindent Put $\displaystyle Z_{h}=\sum_{i=1}^{n}\left( g(X_{i})\mathbb{I}_{]\overset{\_}{u}%
_{h-1}{}_{,}\overset{\_}{u}_{h}]}(X_i)-\mathbb{E}g(X_{i})\mathbb{I}_{]\overset{\_}{u}_{h-1}{}_{,}%
\overset{\_}{u}_{h}]}(X_i)\right) ,$ $h\geq 1.$ \\ We have%
\begin{equation*}
\sqrt{n}(S_{n}(\overset{\_}{u}_{j})-S_{n}(v))=%
\sum_{h=j}^{m(p)}Z_{h}=T_{m(p)-j}
\end{equation*}%
with%
\begin{equation*}
\sqrt{n}(S_{n}(v-\delta
)-S_{n}(v))=\sum_{i=1}^{m(p)}Z_{i}=T_{m(p)}=T(n,u,\delta ).
\end{equation*}

We observe that $\left\{ T_{1},T_{2},...,T_{m(p)}\right\} $ are partial sums
of i.i.d. centered random variables so that the $T_{j}^{4}$ form a
submartingale. By the maximal inequality form submartingales, for any fixed $%
p $

\begin{eqnarray*}
\mathbb{P}(\max_{1\leq j\leq m(p)}\left\vert S_{n}(\overset{\_}{u}_{j})-S_{n}(v)%
\right\vert \geq \eta )=\mathbb{P}(\max_{1\leq j\leq m(p)}\left\vert T_{j}\right\vert 
\geq \eta \sqrt{n})&\leq &\frac{1}{\eta ^{4}n^{2}}\mathbb{E}T_{m(p)}^{4}\\
& \leq &\frac{1}{\eta ^{4}n^{2}}\mathbb{E}T(n,u,\delta )^{4}.
\end{eqnarray*}
\noindent Since the right hand does not depend on $p,$ we get by (\ref%
{approx01})

\bigskip 
\begin{equation*}
\frac{1}{_{\delta }}\mathbb{P}(\sup_{v-\delta \leq u\leq v}\left\vert
S_{n}(u)-S_{n}(v)\right\vert \geq \eta )\leq \frac{1}{\delta \eta ^{4}n^{2}}%
\mathbb{E}T(n,u,\delta )^{4}.
\end{equation*}

\noindent Notice that $T(n,u,\delta )$ is a sum of $n$ i.i.d centered
random variables with variance%
\begin{equation*}
\kappa_1(v,\delta)=\sigma ^{2}(v,\delta )=\int_{v-\delta }^{v}\left( g(x)-\mathbb{E}%
(g(x)\right) ^{2}dF(x)
\end{equation*}%
and fourth moment%
\begin{equation*}
\kappa _{2}(v,\delta )=\int_{v-\delta }^{v}\left( g(x)-\mathbb{E}%
(g(x)\right) ^{4}dF(x).
\end{equation*}
\\
\bigskip
Simple computations give (see the \textsc{appendix} \ref{mocom} for a simple proof of that)%
\begin{equation*}
\mathbb{E}\left( T(n,u,\delta )\right) ^{4}=n\kappa_2(v,\delta
)+3n(n-1)\kappa _{1}^2(v,\delta ).
\end{equation*}%
By putting these facts together, we arrive at %
 \begin{eqnarray*}
\frac{1}{\delta }\mathbb{P}(\sup_{v-\delta \leq u\leq v}\left\vert
S_{n}(u)-S_{n}(v)\right\vert \geq \eta )&\leq& \eta ^{-4}\left( \frac{n\kappa
_{2}(v,\delta )+3n(n-1)\sigma ^{4}(v,\delta )}{\delta n^{2}}\right)\\
&\leq& \eta ^{-4}\left(\frac{\kappa_{2}(v,\delta )}{\delta}\times \frac{1}{n} +3\frac{\sigma ^{4}(v,\delta )}{\delta}\times\Big[1-\frac{1}{n} \Big]\right) .
\end{eqnarray*}
\noindent Remark that 
\begin{eqnarray*}
\sup_{u_{0}\leq v\leq u_{1}}\frac{\kappa _{2}(v,\delta )}{\delta}   &\leq & \left(
\sup_{u_{0}\leq x \leq u_{1}}\left\vert g(x)-\mathbb{E}g(X)\right\vert
\right) ^{4}\times \delta^{-1}\times \sup_{u_{0}\leq v\leq u_{1}}\int_{v-\delta}^v dF(x) \\
&\leq & \left(\sup_{u_{0}\leq x \leq u_{1}}\left\vert g(x)-\mathbb{E}g(X)\right\vert
\right) ^{4}\times \sup_{u_{0}\leq v\leq u_{1}} \frac{F(v)-F(v-\delta)}{\delta}.
\end{eqnarray*}
We finally get

\begin{equation*}
\lim_{\delta \rightarrow 0}\sup_{u_{0}\leq v\leq u_{1}}\sup_{n\geq 1}\frac{1%
}{_{\delta }}\mathbb{P}(\sup_{v-\delta \leq u\leq v}\left\vert
S_{n}(u)-S_{n}(v)\right\vert \geq \eta )=0
\end{equation*}%
whenever $\displaystyle \lim_{\delta \rightarrow 0} \sup_{u_{0}\leq v\leq u_{1}} \frac{\sigma ^{4}(v,\delta )}{\delta}=0$  and $\displaystyle \sup_{u_{0}\leq x\leq
u_{1}}\left\vert g(x)-\mathbb{E}g(X)\right\vert <+\infty .$
\begin{flushright}
$\square$
\end{flushright}
This achieves the proof of the lemma.
\bigskip

\noindent \textbf{Proof of Theorem \ref{theo2}}.

\bigskip
By Theorem 2.7.5 in \cite{vaart} applied to $\mathcal{F}_1$ and by the fact that $\mathcal{F}_2$ is a Vapnik-Chervonenkis class, condition (2.5.1) is satisfied for both $\mathcal{F}_1$ and $\mathcal{F}_2$ thus $\mathcal{F}_1$ and $\mathcal{F}_2$ are Donsker classes.

This may be used in a simple manner to get 
\begin{equation}
A_{n}=\max (\sup_{u \in I}\left\vert \mathbb{G}_{n}(g_{u})\right\vert
,\sup_{u \in I}\left\vert \mathbb{G}_{n}(f_{u})\right\vert )=O_{\mathbb{P%
}}(1,I)\text{ as }n\rightarrow \infty .  \label{anborne}
\end{equation}%
Denote the functional empirical process for any real function $g$ by%
\begin{equation*}
\mathbb{G}_{n}(g)=\frac{1}{\sqrt{n}}\sum_{i=1}^{n}\left\{ g(X_{i})-\mathbb{E}%
g(X_{i})\right\}. 
\end{equation*}%
Remind that for any Donsker class $\mathcal{G}$, the functional stochastic process $\{\mathbb{G}%
_{n}(g),g\in \mathcal{G}\}$ converges in law to a Gaussian and centered stochastic
process $\{\mathbb{G}(g),g\in \mathcal{G}\}$ whose variance-covariance function is%
\begin{equation*}
\Gamma (g_{1},g_{2})=\int (g_{1}(x)-\mathbb{E}g(X_{1})(g_{2}(x)-\mathbb{E}%
g_{2}(X_{1}))dF(x).
\end{equation*}%
We have, as $n\rightarrow \infty $

\begin{equation*}
\mathbb{P}_{n}(g_{u})=\mathbb{P}_{X}(g_{u})+\frac{\mathbb{G}_{n}(g_{u})}{%
\sqrt{n}}
\end{equation*}%
\begin{equation*}
\mathbb{P}_{n}(f_{u})=\mathbb{P}_{X}(f_{u})+\frac{\mathbb{G}_{n}(f_{u})}{%
\sqrt{n}}.
\end{equation*}%
 Thus 
\begin{eqnarray*}
\sqrt{n}({e}_{n}(u)-e(u)) &=&\sqrt{n}\Big(\frac{\mathbb{P}_{n}(f_{u})}{%
\mathbb{P}_{n}(g_{u})}-\frac{\mathbb{P}_{X}(f_{u})}{\mathbb{P}_{X}(g_{u})}%
\Big) \\
&=&\sqrt{n}\Big(\frac{\mathbb{P}_{n}(f_{u})}{\mathbb{P}_{n}(g_{u})}-\frac{%
\mathbb{P}_{X}(f_{u})}{\mathbb{P}_{n}(g_{u})}+\frac{\mathbb{P}_{X}(f_{u})}{%
\mathbb{P}_{n}(g_{u})}-\frac{\mathbb{P}_{X}(f_{u})}{\mathbb{P}_{X}(g_{u})}%
\Big) \\
&=&\displaystyle\frac{1}{\mathbb{P}_{n}(g_{u})}\sqrt{n}\Big(\mathbb{P}%
_{n}(f_{u})-\mathbb{P}_{X}(f_{u})\Big)-\mathbb{P}_{X}(f_{u})\frac{\sqrt{n}%
\Big(\mathbb{P}_{n}(g_{u})-\mathbb{P}_{X}(g(u)\Big)}{\mathbb{P}_{n}(g_{u})%
\mathbb{P}_{X}(g_{u})} \\
&=&\displaystyle\frac{1}{\mathbb{P}_{n}(g_{u})}\Big[\mathbb{G}_{n}(f_{u})-%
\frac{\mathbb{P}_{X}(f_{u})}{\mathbb{P}_{X}(g_{u})}\mathbb{G}_{n}(g_{u}))%
\Big] \\
&=&\displaystyle\frac{1}{\mathbb{P}_{n}(g_{u})}\Big[\mathbb{G}_{n}\Big(f_{u}-%
\frac{\mathbb{P}_{X}(f_{u})}{\mathbb{P}_{X}(g_{u})}g_{u}\Big)\Big].
\end{eqnarray*}%
We find 
\begin{eqnarray*}
(\mathbb{P}_{n}(g_{u}))^{-1} &=&\Big[\mathbb{P}_{X}(g_{u})+\frac{\mathbb{G}%
_{n}(g_{u})}{\sqrt{n}}\Big]^{-1} \\
&=&\mathbb{P}_{X}^{-1}(g_{u})\Big[1+\mathbb{P}_{X}^{-1}(g_{u})\times
n^{-1/2}\times \mathbb{G}_{n}(g_{u})\Big]^{-1} \\
&=&\mathbb{P}_{X}^{-1}(g_{u})\Big[1-\mathbb{P}_{X}^{-1}(g_{u})\times
n^{-1/2}\times \mathbb{G}_{n}(g_{u})+\mathbb{P}_{X}^{-1}(g_{u})\times \theta \Big(%
n^{-1/2}\times \mathbb{G}_{n}(g_{u})\Big)\Big]
\end{eqnarray*}%
Since $\mathcal{F}_{1}$ is a Donsker class, then $\displaystyle\sup_{u\in I
}| \mathbb{G}_{n}(g_{u})| =\Vert \mathbb{G}_{n}\Vert _{\mathcal{F}%
_{1}}=O_{\mathbb{P}}(1,I)$. So
\begin{equation*}
(\mathbb{P}_{n}(g_{u}))^{-1}=\mathbb{P}_{X}^{-1}(g_{u})\Big[1-\mathbb{P}%
_{X}^{-1}(g_{u})\times n^{-1/2}\times O_{\mathbb{P}}(1,I)\Big]
\end{equation*}%
Let us remind that $h_{u}=\mathbb{P}%
_{X}(g_{u})^{-1}f_{u}-\mathbb{P}_{X}(f_{u})\mathbb{P}_{X}^{-2}(g_{u})g_u$. Then, for $u\in I$, we get
\begin{equation*}
\sqrt{n}({e}_{n}(u)-e(u)) =\Big[\mathbb{G}_{n}\Big(f_{u}-\frac{\mathbb{P}%
_{X}(f_{u})}{\mathbb{P}_{X}(g_{u})}g_{u}\Big)\Big]\times \Big[\mathbb{P}%
_{X}^{-1}(g_{u})-\mathbb{P}_{X}^{-2}(g_{u})\times n^{-1/2}\times O_{\mathbb{P%
}}(1,I)\Big]
\end{equation*}

\begin{equation*}
=\mathbb{G}_{n}(h_{u})+\mathbb{G}_{n}(h_{u})\times \mathbb{P}%
_{X}^{-1}(g_{u})\times n^{-1/2}\times O_{\mathbb{P}}(1,I).
\end{equation*}
\\
We finally have 
\begin{equation}
\sqrt{n}({e}_{n}(u)-e(u))=\mathbb{G}_{n}(h_{u})+\mathbb{G}_{n}(h_{u})\times
o_{\mathbb{P}}(1,I). \ \ \ \ \ \ \ \square \label{gn}
\end{equation}

\begin{lemma}
The class $\displaystyle \mathcal{F}_3=\Big\{ h_u=\frac{f_u}{\mathbb{P}_X
(g_{u})} -\frac{\mathbb{P}_X(f_{u})}{\mathbb{P}^{2}_X (g_{u})}g_u, u \in 
I\Big\}$ is a Donsker Class.
\end{lemma}

At this step, we want to prove that $\mathcal{F}_3\mathbb{=\{}h_{u},u_{0}\leq
u\leq u_{1}\}$ is a Donsker Class. Since we obviously have, by the \textit{Central Limit Theorem}, finite distribution
convergence of $\mathbb{\{G}_{n}\mathbb{(}h_{u}),u\in I\}$\
to the stochastic process $\mathbb{\{G(}h_{u}),u\in I\}$ in $%
\ell ^{\infty }(\mathcal{F}_3),$ we only need to prove the asymptotic tightness
of $\{ \mathbb{G}_n(h_{u}),u\in I\}.$ \\
\bigskip
\\ In view of Theorem in
1.5.7 \ in \cite{vaart}, \ it is enough to prove that%
\begin{equation*}
\lim_{\delta \rightarrow 0}\sup_{u\in I}\lim \sup_{n\rightarrow \infty }%
\frac{1}{_{\delta }}\mathbb{P}(\sup_{v-\delta \leq u\leq v}\left\vert
\mathbb{G}_{n}(h_{u})-\mathbb{G}_{n}(h_{v})\right\vert \geq \eta )=0.
\end{equation*}%
Here, we apply \textbf{Lemma \ref{lemg}} for the nondecreasing mesurable function $g(x)=x$ and $g(x)=1$.\\
\bigskip
\\
  In both cases, we inspect the assumptions of this lemma and see that 
if $g(x)=x$, \\ we get  $g(x)\leq g(u_1)=u_1$ for any $ u_0\leq x \leq u_1$ and thus
\begin{eqnarray*}
\sup_{u_0\leq v \leq u_1} \frac{\sigma ^{4}(v,\delta )}{\delta}&=&\sup_{u_0\leq v \leq u_1}  \frac{1}{\delta}\Big( \int_{v-\delta}^v (g(x)-\mathbb{E}g(x))^2 dF(x)\Big)^2  \\
& \leq & \vert u_{1}-\mathbb{E}(X)\vert^4 \times \sup_{u_0\leq v \leq u_1}\Big(\frac{F(v)-F(v-\delta) }{\sqrt{\delta}}\Big)^2  \rightarrow 0 \ \ 
\text{as }\delta \rightarrow 0 ,
\end{eqnarray*}
and%
\begin{equation*}
\sup_{x\in I}\left\vert g(x)-\mathbb{E}g(X)\right \vert    \leq \left\vert
u_{1} \vert+ \vert \mathbb{E}(X)\right\vert<\infty .
\end{equation*}
If $g(x)=1$, the result is obvious. \\
\bigskip
\\
We can 
 apply \textbf{Lemma \ref{lemg}} and we will get$,$%
\begin{equation*}
\lim_{\delta \rightarrow 0}\sup_{u\in I}\limsup_{n\rightarrow \infty }%
\frac{1}{_{\delta }}\mathbb{P}(\sup_{v-\delta \leq u\leq v}\left\vert
\mathbb{G}_{n}(f_{u})-\mathbb{G}_{n}(f_{v})\right\vert \geq \eta )=0
\end{equation*}%
and%
\begin{equation*}
\lim_{\delta \rightarrow 0}\sup_{u\in I}\limsup_{n\rightarrow \infty }%
\frac{1}{_{\delta }}\mathbb{P}(\sup_{v-\delta \leq u\leq v}\left\vert
\mathbb{G}_{n}(g_{u})-\mathbb{G}_{n}(g_{v})\right\vert \geq \eta )=0.
\end{equation*}%
But by Theorem 8.3 of Billingsley \cite{bil}, p.56, and by Theorem 2.2 in Lo 
\cite{loasymp}, these two previous equalities 
 entail
, that \begin{equation*}
\lim_{\delta \rightarrow 0}\sup_{u\in I}\limsup_{n\rightarrow \infty }%
\mathbb{P}(\sup_{\left\vert u-v\right\vert \leq \delta ,(u,v)\in
I^{2}}\left\vert \mathbb{G}_{n}(f_{u})-\mathbb{G}_{n}(f_{v})\right\vert \geq \eta )=0
\end{equation*}%
and%
\begin{equation*}
\lim_{\delta \rightarrow 0}\sup_{u\in I}\limsup_{n\rightarrow \infty }%
\mathbb{P}(\sup_{\left\vert u-v\right\vert \leq \delta ,(u,v)\in
I^{2}}\left\vert \mathbb{G}_{n}(g_{u})-\mathbb{G}_{n}(g_{v})\right\vert \geq \eta )=0.
\end{equation*}%
Next, we use the following development for $(u,v)\in I^{2}$ 
\begin{eqnarray*}
h_{u}-h_{v} &=&\mathbb{P}_{X}^{-1}(g_{u})\Big(f_{u}-\frac{\mathbb{P}%
_{X}(f_{u})}{\mathbb{P}_{X}(g_{u})}g_{u}\Big)-\mathbb{P}_{X}^{-1}(g_{v})\Big(%
f_{v}-\frac{\mathbb{P}_{X}(f_{v})}{\mathbb{P}_{X}(g_{v})}g_{v}\Big) \\
&=&\underbrace{\mathbb{P}_{X}^{-1}(g_{u})f_{u}-\mathbb{P}%
_{X}^{-1}(g_{v})f_{v}}-\Big(\underbrace{\frac{\mathbb{P}_{X}(f_{u})}{\mathbb{%
P}_{X}^{2}(g_{u})}g_{u}-\frac{\mathbb{P}_{X}(f_{v})}{\mathbb{P}%
_{X}^{2}(g_{v})}g_{v}}\Big) \\
&=&\ \ \ \ \ \ \ \ \ \ \ a(u,v)\ \ \ \ \ \ \ \ \ \ -\ \ \ \ \ \ \ \ \ \ \ \
b(u,v)
\end{eqnarray*}%
We get%
\begin{eqnarray*}
a(u,v) &=&\mathbb{P}_{X}^{-1}(g_{u})f_{u}-\mathbb{P}_{X}^{-1}(g_{u})f_{v}+%
\mathbb{P}_{X}^{-1}(g_{u})f_{v}-\mathbb{P}_{X}^{-1}(g_{v})f_{v} \\
&=&(f_{u}-f_{v})\times \mathbb{P}_{X}^{-1}(g_{u})+\Big(\mathbb{P}%
_{X}^{-1}(g_{u})-\mathbb{P}_{X}^{-1}(g_{v})\Big)\times f_{v} \\
&=&
\dfrac{f_{u}-f_{v}}{\mathbb{P}_{X}(g_{u})}-\frac{\mathbb{P}%
_{X}(g_{u})-\mathbb{P}_{X}(g_{v})}{\mathbb{P}_{X}(g_{u})\times \mathbb{P}%
_{X}(g_{v})}\times f_{v}.
\end{eqnarray*}%
Then 
\begin{equation*}
|\mathbb{G}_{n}(a(u,v))|\leq \dfrac{1}{\mathbb{P}_{X}(g_{u})}\times |\mathbb{%
G}_{n}(f_{u}-f_{v})|+\frac{|\mathbb{P}_{X}(g_{u})-\mathbb{P}_{X}(g_{v})|}{%
\mathbb{P}_{X}(g_{u})\times \mathbb{P}_{X}(g_{v})}\times \mathbb{G}%
_{n}(f_{v}).
\end{equation*}%
Next%
\begin{eqnarray*}
b(u,v) &=&\frac{\mathbb{P}_{X}(f_{u})}{\mathbb{P}_{X}^{2}(g_{u})}\times
g_{u}-\frac{\mathbb{P}_{X}(f_{u})}{\mathbb{P}_{X}^{2}(g_{u})}\times g_{v}+%
\frac{\mathbb{P}_{X}(f_{u})}{\mathbb{P}_{X}^{2}(g_{u})}\times g_{v}-\frac{%
\mathbb{P}_{X}(f_{v})}{\mathbb{P}_{X}^{2}(g_{v})}\times g_{v} \\
&=&(g_{u}-g_{v})\times \frac{\mathbb{P}_{X}(f_{u})}{\mathbb{P}_{X}^{2}(g_{u})%
}+\Big[\frac{\mathbb{P}_{X}(f_{u})}{\mathbb{P}_{X}^{2}(g_{u})}-\frac{\mathbb{%
P}_{X}(f_{v})}{\mathbb{P}_{X}^{2}(g_{v})}\Big]\times g_{v}.
\end{eqnarray*}%
\\
\bigskip
\\
Next,
\bigskip
\\
\begin{eqnarray*}
\frac{\mathbb{P}_{X}(f_{u})}{\mathbb{P}_{X}^{2}(g_{u})}-\frac{\mathbb{P}%
_{X}(f_{v})}{\mathbb{P}_{X}^{2}(g_{v})} &=&\frac{\mathbb{P}_{X}(f_{u})}{%
\mathbb{P}_{X}^{2}(g_{u})}-\frac{\mathbb{P}_{X}(f_{u})}{\mathbb{P}%
_{X}^{2}(g_{v})}+\frac{\mathbb{P}_{X}(f_{u})}{\mathbb{P}_{X}^{2}(g_{v})}-%
\frac{\mathbb{P}_{X}(f_{v})}{\mathbb{P}_{X}^{2}(g_{v})} \\
&=&\Big(\frac{1}{\mathbb{P}_{X}^{2}(g_{u})}-\frac{1}{\mathbb{P}%
_{X}^{2}(g_{v})}\Big)\times \mathbb{P}_{X}(f_{u}) \\
&& \ \ \ \ \   \ \ \   \ \ \ \ \  + \ \ \ \Big(\mathbb{P}_{X}(f_{u})-\mathbb{P}_{X}(f_{v})\Big)\times \frac{1}{%
\mathbb{P}_{X}^{2}(g_{v})} \\
&=&\Big(\frac{(\mathbb{P}_{X}(g_{v})+\mathbb{P}_{X}(g_{u}))\times (\mathbb{P}%
_{X}(g_{v})-\mathbb{P}_{X}(g_{u}))}{\mathbb{P}_{X}^{2}(g_{u})\times \mathbb{P%
}_{X}^{2}(g_{v})}\Big)\times \mathbb{P}_{X}(f_{u}) \\
&& \ \ \ \ \   \ \ \   \ \ \ \ \  + \ \ \ \Big(\mathbb{P}_{X}(f_{u})-\mathbb{P}_{X}(f_{v})\Big)\times \frac{1}{%
\mathbb{P}_{X}^{2}(g_{v})}.
\end{eqnarray*}%
\bigskip
\\
Also,
\begin{eqnarray*}
|\mathbb{G}_{n}(b(u,v))| &\leq &\dfrac{|\mathbb{P}_{X}(f_{u})|}{\mathbb{P}%
_{X}^{2}(g_{u})}\times |\mathbb{G}_{n}(g_{u}-g_{v})| \\
&& \ \ \ \ \  + \ \ \ \frac{|\mathbb{P}_{X}(g_{v})+\mathbb{P}_{X}(g_{u})|\times |\mathbb{P}%
_{X}(g_{v})-\mathbb{P}_{X}(g_{u}))}{\mathbb{P}_{X}^{2}(g_{u})\times \mathbb{P%
}_{X}^{2}(g_{v})|}\times |\mathbb{P}_{X}(f_{u})|\times |\mathbb{G}%
_{n}(g_{v})| \\
&& \ \ \ \ \   \ \ \ \ \ \ \ \   \ \ \   \ \ \ \ \   \ \ \ \ \ \ \ \ \ \   \ \ \  \ \ \ \ \ + \ \ \ |\mathbb{P}_{X}(f_{u})-\mathbb{P}_{X}(f_{v})|\times |\mathbb{G}%
_{n}(g_{v})|\times \frac{1}{\mathbb{P}_{X}^{2}(g_{v})}.
\end{eqnarray*}
\bigskip
\\
For $(u,v)\in \lbrack u_{0},u_{1}]^2$,
let us use the bounds of $\mathbb{P}_{X}^{-1}(g_{u})$, $\mathbb{P}_{X}^{-1}(g_{v})$, and $\mathbb{P}_{X}(f_{u}).$ \\
\bigskip
\\We obtain $\mathbb{P}_X^{-1}(g_u)\leq (\bar{F}(u_1))^{-1}$, $\mathbb{P}_X^{-1}(g_v)\leq (\bar{F}(u_1))^{-1}$, and finally, from \eqref{fuesp}, we get $\vert \mathbb{P}_{X}(f_{u})\vert \leq \mathbb{E}\vert X \vert.$ 
\\
\bigskip
\\
 Thus by using these bounds and (\ref{anborne})$,$ it comes that%
\begin{equation*}
\sup_{\left\vert u-v\right\vert \leq \delta ,(u,v)\in I^{2}}\left\vert 
\mathbb{G}_{n}(h_{u}-h_{v})\right\vert \leq B_{1}\times \sup_{\left\vert
u-v\right\vert \leq \delta ,(u,v)\in I^{2}}\left\vert \mathbb{G}%
_{n}(f_{u}-f_{v})\right\vert + \, B_{2}\times \sup_{\left\vert u-v\right\vert \leq
\delta ,(u,v)\in I^{2}}\left\vert \mathbb{G}_{n}(g_{u}-g_{v})\right\vert
\end{equation*}%
\begin{equation*}
+\left( B_{3} \times \sup_{\left\vert u-v\right\vert \leq \delta ,(u,v)\in
I^{2}}\left\vert \mathbb{P}_{X}(g_{u})-\mathbb{P}_{X}(g_{v})\right\vert
+B_{4} \times\sup_{\left\vert u-v\right\vert \leq \delta ,(u,v)\in I^{2}}\left\vert 
\mathbb{P}_{X}(f_{u})-\mathbb{P}_{X}(f_{v})\right\vert \right) A_{n},
\end{equation*}
\bigskip
where $$ \begin{cases}
B_1=(\bar{F}(u_1))^{-1};\\
B_2=\mathbb{E}\vert X \vert \times (\bar{F}(u_1))^{-2};\\
B_3=\bar{F}(u_1))^{-2}\Big((\bar{F}(u_1))^{-2}+2\bar{F}(u_0)\Big);\\
B_4=(\bar{F}(u_1))^{-2};\\
\displaystyle A_{n}=\max (\sup_{u \in I}\left\vert \mathbb{G}_{n}(g_{u})\right\vert
,\sup_{u \in I}\left\vert \mathbb{G}_{n}(f_{u})\right\vert ).
\end{cases}$$
\bigskip\\

Now we observe that 
\begin{equation*}
\sup_{\left\vert u-v\right\vert \leq \delta ,(u,v)\in I^{2}}\left\vert 
\mathbb{P}_{X}(g_{u})-\mathbb{P}_{X}(g_{v})\right\vert = \sup_{\left\vert
u-v\right\vert \leq \delta ,(u,v)\in I^{2}}\left\vert F(u)-F(v)\right\vert
\end{equation*}%
and%
\begin{eqnarray*}
\sup_{\left\vert u-v\right\vert \leq \delta ,(u,v)\in I^{2}}\left\vert 
\mathbb{P}_{X}(f_{u})-\mathbb{P}_{X}(f_{v})\right\vert &\leq& \sup_{\left\vert u-v\right\vert \leq \delta ,(u,v)\in I^{2}} \left \vert \int_u^\infty t dF(t) - \int_v^\infty tdF(t)  \right \vert\\
& \leq &
 \sup_{\left\vert u-v\right\vert \leq \delta ,(u,v)\in I^{2}}\left\vert \int_u^v tdF(t) \right\vert\\
& \leq &
\max(\vert u_{0} \vert,\vert u_{1} \vert) \sup_{\left\vert u-v\right\vert \leq \delta ,(u,v)\in I^{2}}\left\vert
F(u)-F(v)\right\vert.  
\end{eqnarray*}
These quantities go to zero whenever $F$ is continuous and hence
uniformly continuous in $I.$  Putting all these facts together and using (\ref{anborne})
yield 
\begin{equation*}
\sup_{n\geq 1}\sup_{\left\vert u-v\right\vert \leq \delta ,(u,v)\in
I^{2}}\left\vert \mathbb{G}_{n}(h_{u}-h_{v})\right\vert \rightarrow 0 \ \ \text{
as }\ \ \delta \rightarrow 0.
\end{equation*}%
Finally $\mathcal{F}_3$ is a Donsker class, thus $\displaystyle \sup_{u\in I}\left\vert \mathbb{G}_{n}(h_{u})\right\vert =O_{\mathbb{P}}(1,I)$ and  we get from (\ref{gn}) that 
\begin{equation*}
\sqrt{n}({e}_{n}(u)-e(u))=\mathbb{G}_{n}(h_{u})+o_{\mathbb{P}}(1,I).
\end{equation*}\begin{flushright}
$\square$
\end{flushright}
This completes the proof.\\
\bigskip
\\
Now we are going to concentrate on consistency bands for the mean excess function.
\section{Consistency bands\label{sec4}}
Now, we may use the uniform bands of the functional empirical processes based
on Talagrand's inequality (see \cite{talagrand}) and the new methods introduced
by Mason and al. \cite{EM2010} to obtain consistency bands of the mean
excess function as follows.

\bigskip

\begin{theorem}
\label{theo3} Let $X_{1},$ $X_{2},\cdots $, be i.i.d random variables with
finite second moment. Put $I=[u_{0},u_{1}],$ with $-\infty <
u_{0}<u_{1}<x_{F}$.
\\
We suppose that $F$ is continuous and satisfies%
\begin{equation}\limsup_{\delta \rightarrow 0} \sup_{\left( v,v-\delta \right) \in I^{2}} 
\left( \frac{ F(v)-F(v-\delta )}{\sqrt{\delta}}\right)^2 =0.
  \label{Fdelta}
\end{equation}
Then for any $\varepsilon >0,$ there exists $n_{0}$ such that for $n\geq
n_{0},$

\begin{equation*}
\mathbb{P}\Big(e_{n}(u)-\frac{E_{n}}{\sqrt{n}}<e(u)<e_n(u)+\frac{E_{n}}{%
\sqrt{n}},u\in I\Big)\geq 1-\varepsilon,
\end{equation*}
with 
\begin{equation}E_{n}=\frac{1}{\overline{F}(u_{1})-D_{1}/\sqrt{n}}\Big(D_{2}+ \frac{D_{1}\times\mathbb{E}\vert X \vert}{ 
\overline{F}(u_{1})}\Big),  \label{enband}
\end{equation}
and where
\begin{equation*}
\left\{ 
\begin{array}{l}
\displaystyle D_{1}= 2 A A_1 \sqrt{\log 2}+A_1 \cr \displaystyle D_{2}= A\,A_1\,M_{1} \sqrt{\log M_{1}}+A_1 ,\cr%
M_1=\max (2 ,\max(\vert u_0 \vert,\vert u_1 \vert))\cr
\end{array}%
\right.
\end{equation*} 
 $A$ and $A_1$ are universal constants.
\end{theorem}
 
The proof of this theorem is rather technical so we postpone it in the
\textsc{appendix } \textsc{subsection} \ref{unifasc} where we also state 
 the fundamental Talagrand's inequality.
\newpage
\textbf{Remark}: The validity condition \eqref{Fdelta} is quite very weak
and is satisfied by most of the continuous usual distribution functions.
Indeed if $F$ is absolutely continuous with respect to the Lebesgue measure
with derivative function $f$, we get by using the 
 \textit{mean value theorem},%
  $$\left( \frac{ F(v)-F(v-\delta )}{\sqrt{\delta}}\right)^2\leq \delta\times\sup_{x \in [v-\delta,v]%
}f^2(x). $$
But $\displaystyle \sup_{x \in [v-\delta,v]}f^2(x)<\infty$ whenever $f$ is continuous, by a
simple argument from real analysis. This allows consistency bands for a
huge number of absolutely continuous distribution functions. All the examples
in the   \textsc{Section} \ref{sec5} are devoted to simulations satisfy \eqref{Fdelta}
through this argument.

\bigskip

Now, we are going to focus on the applications of our results.

\section{Simulations and applications \label{sec5}}
\subsection{Introduction}

The Mean excess function can be used in two ways :

\bigskip

$\bullet $ First, it can be used to distinguish heavy tailed models
distribution and those with light tailed distribution. An increasing mean
excess function $e(u)$ indicates  a heavy-tailed distribution and a
decreasing mean excess function $e(u)$ indicates a light-tailed
distribution. The exponential distribution has a constant mean excess
function and is considered a medium-tailed distribution . 

\bigskip

Then the plot of the mean excess function tends to infinity for heavy-tailed
distributions%
, decreases to zero for light-tailed distributions and remains constant for
an exponential distribution.\\

$\bullet $ Secondly, it can be used for tail estimation with the help of the
generalized Pareto distribution 
which can 
model the tails of another distribution.
\\
Let $F_{u}(x)$, the excess distribution over threshold $u$, defined by 
\begin{equation*}
F_{u}(x)=\mathbb{P}(X-u\leq x|X>u)
\end{equation*}%
with $0\leq x<x_{F}-u$, where $x_{F}\leq \infty $ is the right endpoint of $F
$.
\\
\bigskip
\\
By using Theorem 7.20 in \cite{PiBaHaan}
, a natural approximation of $F_{u}$ is a
generalized Pareto distribution $\mathcal{G}PD(\xi ,\beta )$ which mean
excess function is given by  
\begin{equation}
\displaystyle e(u)=\frac{\beta }{1-\xi }+\frac{\xi }{1-\xi }u, \ \ \text{provided that} \ \  \xi <1. \label{gpd1}
\end{equation}
If the empirical mean excess function plot looks linear, we can fit a $\mathcal{G}PD(\xi ,\beta )$ model whose parameters can be estimated by means of linear least squares method : given data $\{(u_1,y_1),\ldots,(u_n,y_n)\},$ where $u_i=X_i $ and $y_i=e_n(u_i),\,i=1,\ldots,n,$ we estimate the parameters $\xi 
$ and $\beta $ to be 
\begin{equation*}
 \hat{\xi}=\frac{\hat{a}}{\hat{a}+1}\ \ \mbox{and}\ \ \hat{\beta}=\frac{\hat{b%
}}{\hat{a}+1},
\end{equation*}%
where 

 $$ \hat{a}=\frac{\displaystyle n\sum_{i=1}^n u_iy_i- \sum_{i=1}^n u_i\sum_{i=1}^n y_i }{\displaystyle n\sum_{i=1}^nu_i^2-\Big(\sum_{i=1}^nu_i\Big)^2}\ \ \mbox{and}\ \ \hat{b}=\overline{y}-\hat{a}\overline{u},$$ with   $\displaystyle \overline{u}=\frac{1}{n}\sum_{i=1}^n u_i$ and $\displaystyle \overline{y}=\frac{1}{n}\sum_{i=1}^n y_i$
are the sample means of the observations on $u$ and $y$, respectively.
\\
\bigskip 
\\
As far as we are concerned, our goal is to estimate the mean excess function by consistency bounds.
\\ 
\bigskip 
\\
In the remainder of this section, we are backing on the empirical mean excess function (\textit{emef} for short) to construct graphical tools goodness of fit test.
\\
\bigskip
\\
In the first step we are considering a large set of distributions for which we draw the average \textit{emef}. That means that we fix a distribution function and consider $n=6000$ samples from it, each sample size is  $4000$. Next we compute the average of the $n=6000$ empirical mean functions.
\\
\bigskip
\\
The graphs of these average mean empirical functions would serve as stallions in the following sense: each other sample having an alike \textit{emef} will suggest such an underlying distribution.  
\\
\bigskip
\\
We will use, as a special guest, the generalized hyperbolic (\textit{Gh} for short) family of distribution functions. Nowadays, this family is very important in financial modeling.\\
 \bigskip
 \\
In a second step we will try to use the obtained graphs as stallions for real data.\\
In this paper, we focus on monthly returns and log-returns of Dow Jones data. We will see that these data strongly suggest \textit{Gh} model.\\
\bigskip
\\
This section, beyong financial data, shows how to use the \textit{emef} for goodness of fit testing purposes. It opens a great verity of applications for differents types of data.
\subsection{Usual distributions}
To assess the performance of our estimator, we present a simulation study. 
We draw simulated \textit{emefs} for standard distributions and next for \textit{Gh} family of distribution functions
\subsubsection{\textit{Emef} for standard distributions.}
We consider some simple models that are listed in the \textsc{table}  \ref{stdist} below where the used parameters are specified and the \textit{emef} figures corresponding to each choice are displayed.
\\  
\begin{center}
\begin{tabular}{|l|l|l|}
\hline 
Distributions & Parameters & Figures \\ 
\hline 
\multirow{2}{*}{GPD}  & $\xi=0.25$, $\beta=1$ &  \\ 
\cline{2-2}
&$\xi=-0.75$, $\beta=1$ &  \multirow{-2}*{Figure \ref{gpd} }\\
\hline
Pareto & $\alpha=7,$ $\lambda=3$ &  \\ 
\cline{2-2} 
Exponential & $\lambda=2$ &  \multirow{-2}*{Figure \ref{pe} } \\ 
\hline 
\multirow{2}{*}{Weibull}  & $\beta=1,\,\tau=3.6$ &  \\ 
\cline{2-2}
& $\beta=1.5,\,\tau=0.2 $ &  \multirow{-2}*{Figure   \ref{weib} }\\
\hline
Burr & $\alpha=0.5,\,\lambda=0.5,\,\tau=5$ &  \\ 
Gomberz & $\alpha=1,\,\lambda=0.5$ & \multirow{-2}*{Figure  \ref{burgom} }\\ 
\hline 
  
Gamma & $\alpha =2,\,\beta =0.001$ &  \\ 
\cline{2-2}
 Beta & $\lambda=7,\beta=2$ & \multirow{-2}*{Figure \ref{gabe}} \\ 
\hline
Lognormal & $ \mu=0, \sigma=1 $ &  \\
\cline{2-2}
 Normal & $ \mu=0, \sigma=1 $ &  \multirow{-2}*{Figure \ref{lono}}\\
\hline 
Laplace & $ \mu=0,\, \sigma=1,\,\tau=0.5$ 
&  Figure \ref{lap1} \\
\hline
$t$ Student  & $ \nu=5,\mu=1$ & \\
\cline{2-2}
Cauchy  & $ \mu=0$, $\delta=1$ &  \multirow{-2}*{Figure \ref{stuchy} }\\
\hline
\end{tabular}
\vspace{3ex}
 \captionof{table}{The \textit{ emef} for standard distributions\label{stdist}}
\end{center}
 \begin{figure}
\includegraphics[scale=0.3]{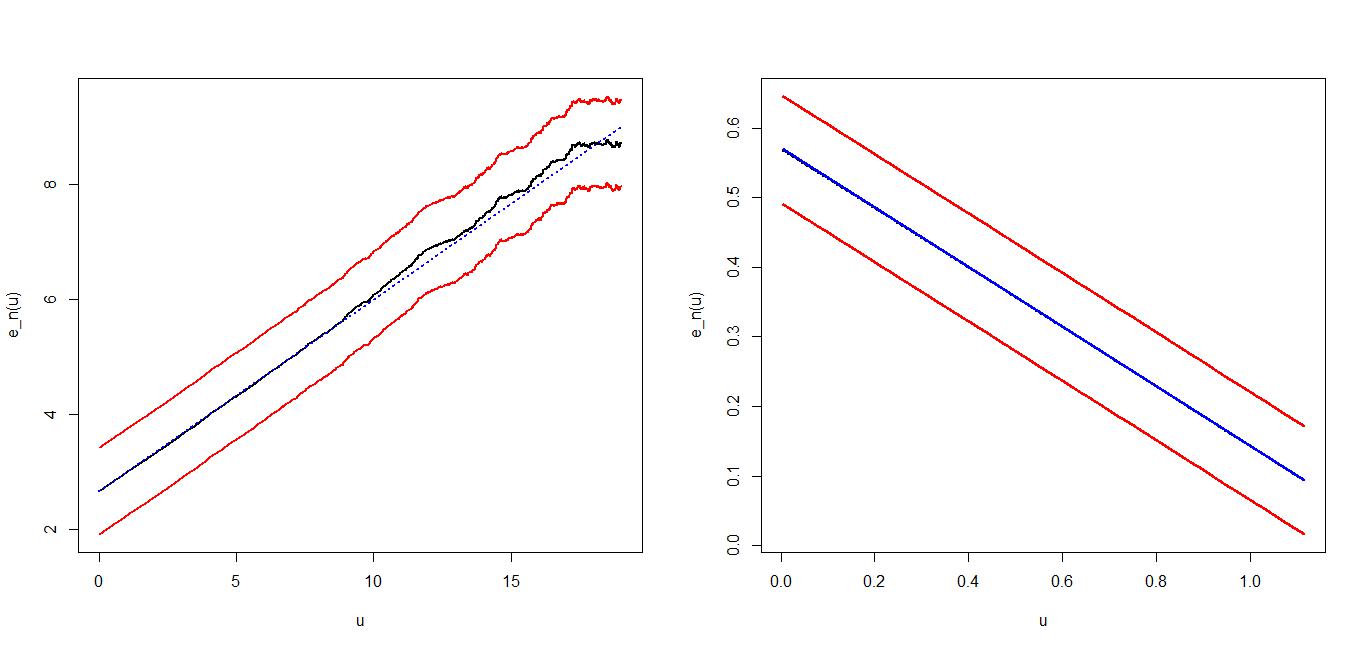} 
\caption{The \textit{emef} for two generalized Pareto distributions : the left panel concerns the one with the parameters $\xi=0.25,\,\beta=1$ and the right panel concerns the one with the parameters $\xi=-0.75,\, \beta=1$.}\label{gpd}
\end{figure}

\begin{figure}
\includegraphics[scale=0.3]{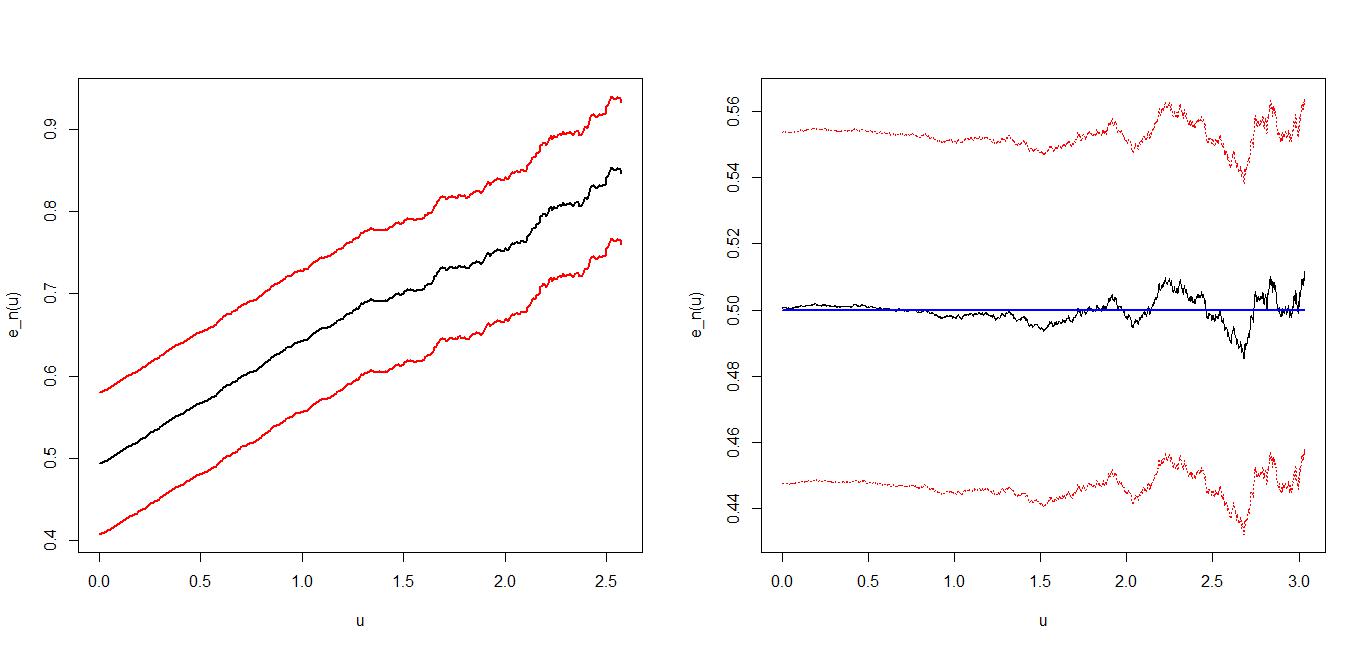} 
\caption{The left panel is the \textit{emef} for a Pareto distribution with the parameters $\alpha=7$ and $\lambda=3$  and the right panel is the one for an Exponential distribution with the parameter $\lambda=2$.}\label{pe}
\end{figure}

\begin{figure}
\includegraphics[scale=0.3]{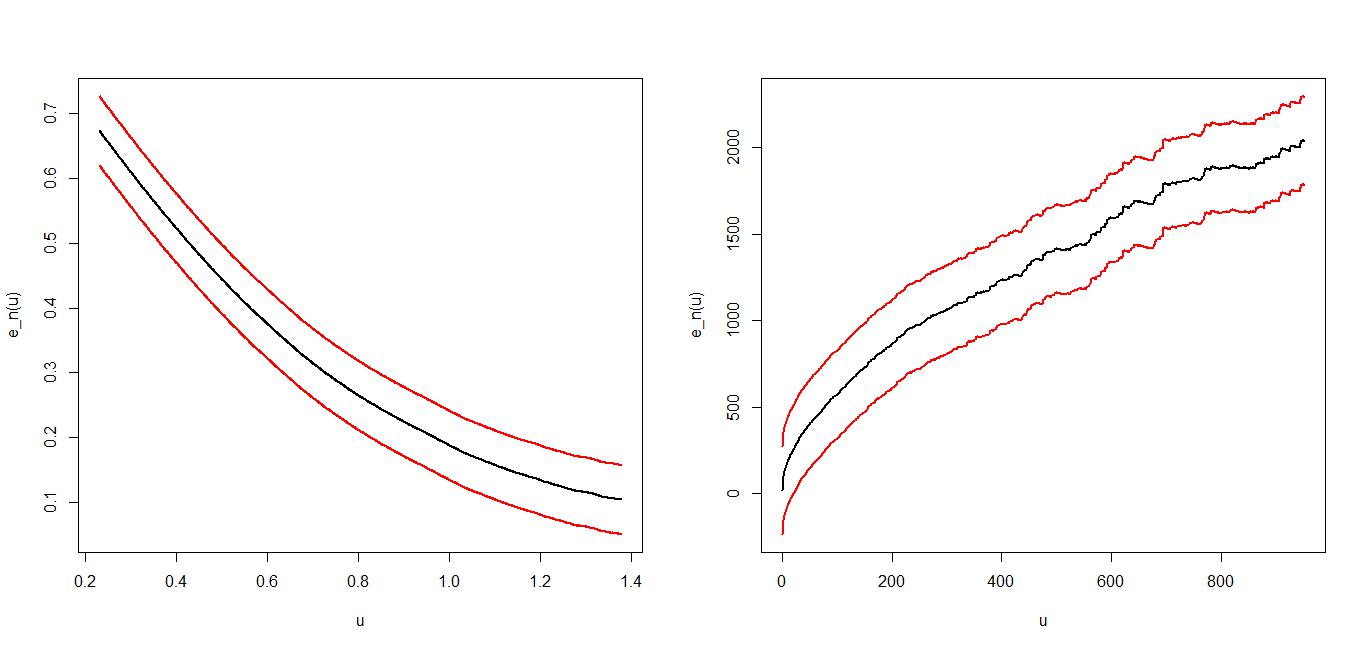}
\caption{The \textit{emef} for two Weibull distributions. The left panel concerns the one with the parameters $\beta=1,\,\tau=3.6$ and the right panel concerns the one with the parameters \\ $\beta=1.5,\,\tau=0.2$.}\label{weib} 
\end{figure}

\begin{figure}
\includegraphics[scale=0.3]{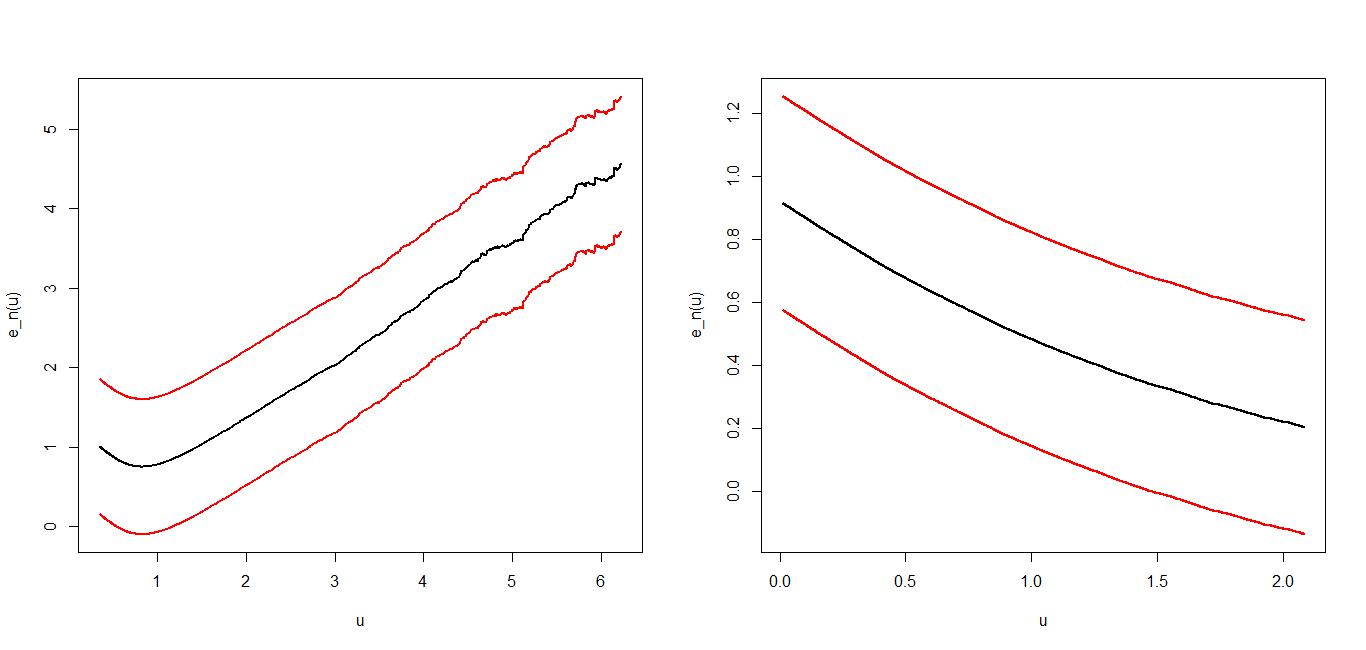}
\caption{The left panel is the \textit{emef} for the Burr distribution with the parameters  $\alpha=0.5,\,\\ \lambda=0.5,\,\tau=5$ and the right one is the \textit{emef} for the Gomberz distribution with the parameters $\alpha=1,\,\lambda=0.5$.} \label{burgom}
\end{figure} 
\begin{figure}
\includegraphics[scale=0.3]{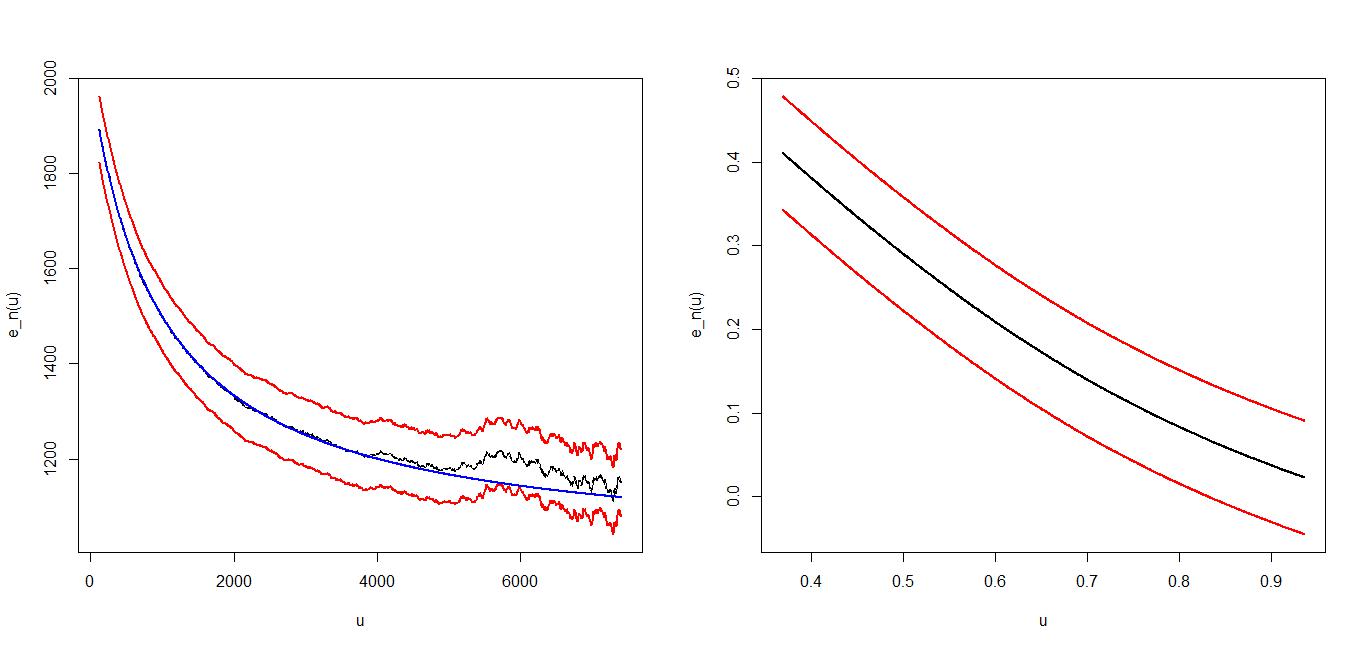} 
\caption{The left panel is the \textit{emef} for the Gamma distribution with the parameters $\alpha =2,\,\beta =0.001$ and the right one is the \textit{emef} for the Beta distribution with the parameters $\lambda=7,\beta=2$.}\label{gabe}
\end{figure}

\begin{figure}
\includegraphics[scale=0.3]{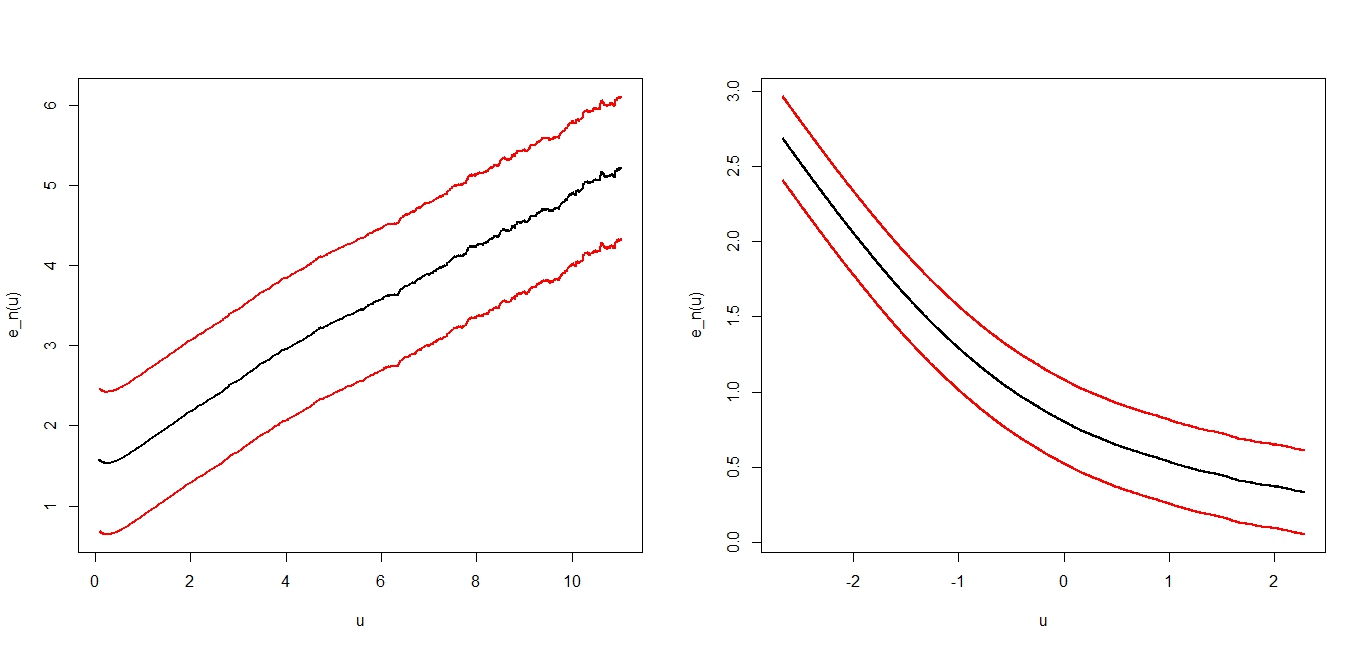} 
\caption{The left panel is the \textit{emef} for the Lognormal distribution with mean $\mu=0$ and with variance  $\sigma^2=1$ and the right one is the \textit{emef} for the Normal distribution with the same parameters. }\label{lono}
\end{figure}

\begin{figure}
\includegraphics[scale=0.3]{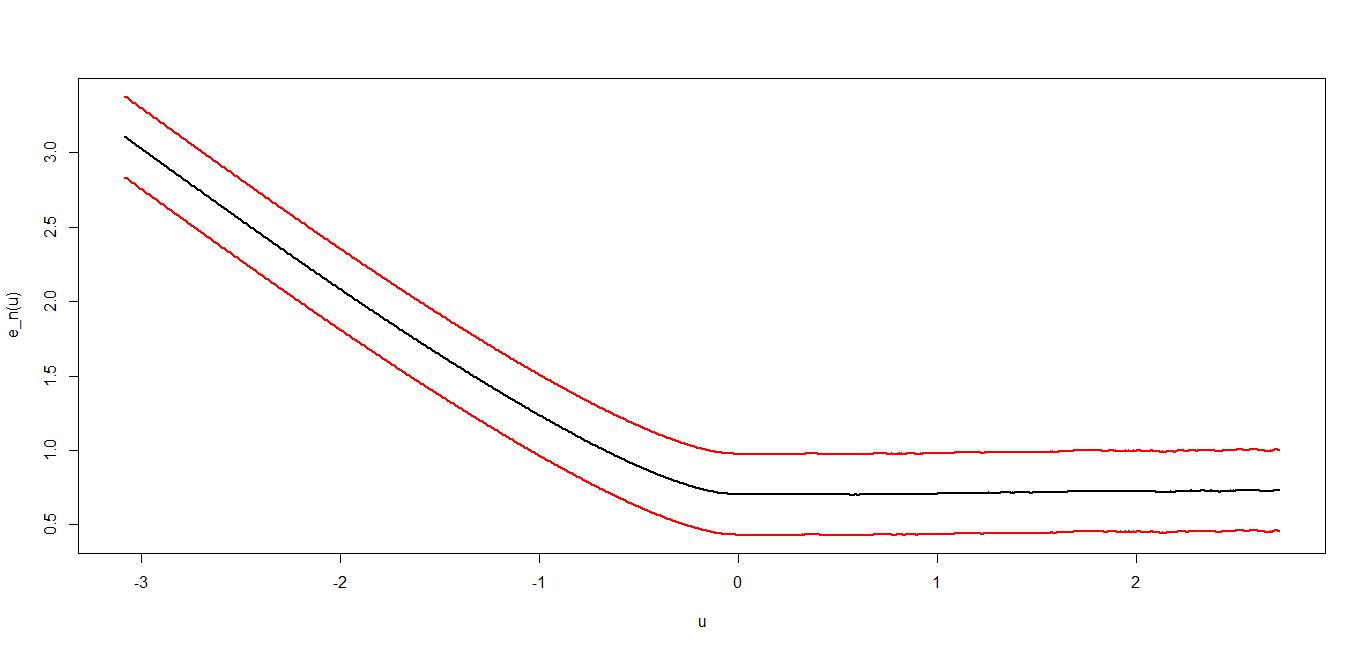}
\caption{\textit{Emef} for Laplace distribution with the parameters $ \mu=0,  \sigma=1, \,\tau=0.5$ }\label{lap1}  
\end{figure}

\begin{figure}
\includegraphics[scale=0.3]{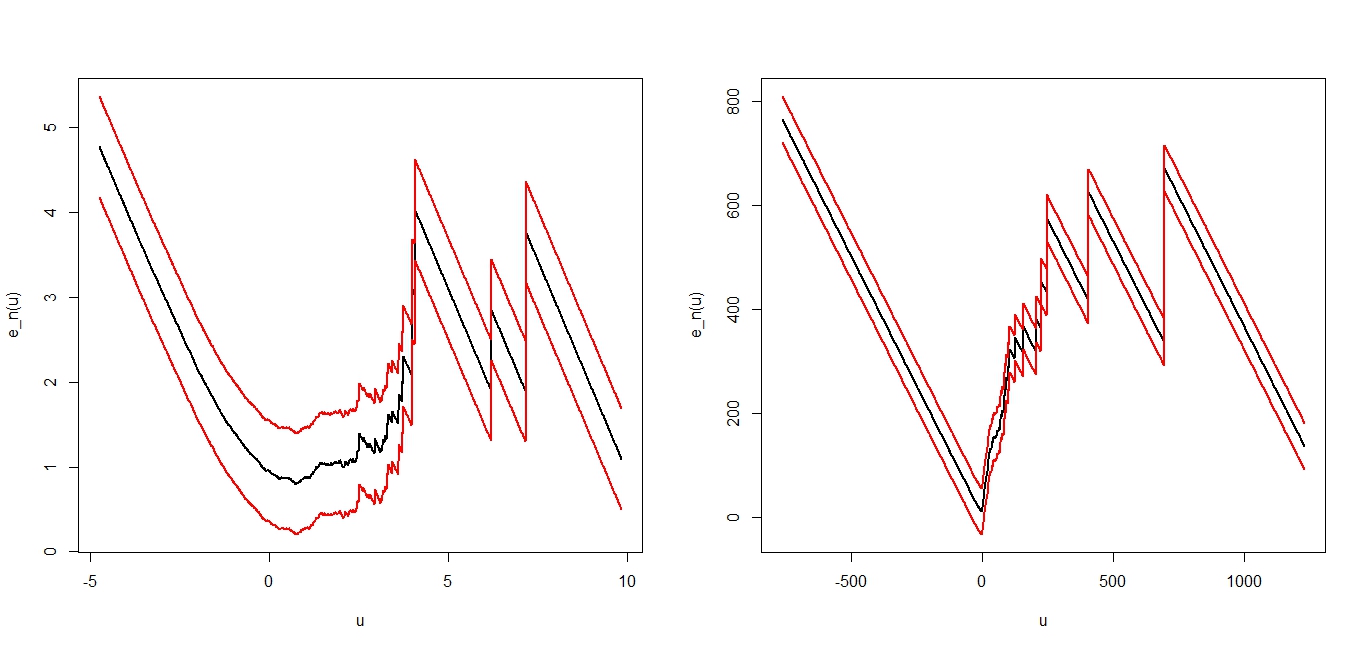}
\caption{\textit{Emef} for a $t$ Student with  $\nu=5$ degrees of freedom and  skewing parameter $\mu=0$ (left panel) and for a Cauchy distribution with location  parameter $\mu=0$ and scale parameter $\delta=1$  (right panel).} \label{stuchy}
\end{figure}

\newpage

\subsubsection{Generalized hyperbolic models}Next, we consider the \textit{emefs} for the \textit{Gh} models. We need some definitions.
The Lebesgue density function of the one dimensional \textit{Gh} is given by \begin{eqnarray}
 f_{\lambda,\alpha, \beta,\delta,\mu}(x)&=&\textbf{a}_{(\lambda,\alpha, \beta,\delta,\mu)}\times \Big(\delta^2+(x-\mu)^2 \Big)^{(\lambda-\frac{1}{2})/2}e^{\beta(x-u)}\times  K_{\lambda-\frac{1}{2}}(\alpha \sqrt{\delta^2+(x-\mu)^2 })
\end{eqnarray}

where  
$$\displaystyle \textbf{a}_{(\lambda,\alpha, \beta,\delta,\mu)}=\frac{(\alpha^2-\beta^2)^{\frac{\lambda}{2}}}{\sqrt{2\pi} \alpha^{(\lambda-\frac{1}{2})}\delta^\lambda K_\lambda (\delta\sqrt{\alpha^2-\beta^2})}$$ is a norming constant to make the curve area equal to $1$ and 
$$\displaystyle K_\lambda(x)=\frac{1}{2}\int_0^\infty y^{\lambda -1}\exp\Big(-\frac{1}{2}x(y+y^{-1}) \Big)dy, \ \  (x>0) $$
is the modified  Bessel function of the third kind with index $\lambda$.\\
\bigskip
\\
The dependence of the parameters $\lambda,\alpha,\beta,\delta,$ and $\mu$ is as follows: $\alpha>0$ determines the \textit{shape}, $0\leq \vert \beta \vert <\alpha$ the \textit{skewness}, $\mu \in \mathbb{R}$ is a \textit{location} parameter and $ \delta>0$ serves for
\textit{scaling}. The parameter $\lambda \in \mathbb{R}$  specifies the order $K_{\lambda}$ function Bessel  that appears in the \textit{Gh} density function  and is used to obtain different subclasses of \textit{Gh} distribution.\\
\bigskip
\\
In the following, we summarise the differents domains of possibilities for the parameters 
 \begin{eqnarray*}
 &&\mbox{If} \ \  \lambda<0 , \ \ \big( \delta>0, \ \ \,  \vert \beta \vert \leq \alpha \big), \\
 && \mbox{if}\ \  \lambda=0 ,\ \ \big( \delta>0, \ \  \,  \vert \beta \vert<\alpha \big), \\
 && \mbox{if}\ \  \lambda> 0, \ \ \big( \delta\geq 0, \ \  \,  \vert \beta \vert<\alpha \big).
\end{eqnarray*}
\\
\bigskip

An important \textit{Gh} family aspect is that it embraces many special cases such that Hyperbolic ($\lambda=1$), 
Student-$t\, (\lambda<0$), Variance Gamma ($\lambda>0$), and the Normal Inverse Gaussian (NIG) ($\lambda=-0.5$) distributions.\\ 
\bigskip
\\
It nests also Generalized Inverse Gamma (GIG) distribution defined only by the three parameters $\lambda, \alpha,$ and $\beta$. An Inverse Gaussian (IG) distribution is a GIG distribution with $\lambda=-0.5$ and a Gamma ($\Gamma$) distribution is also a GIG distribution with $\beta=0$. \\
\bigskip

It contains some limiting distributions such as Cauchy distribution with parameters $\mu$ and $\delta$ (obtained for $\lambda=-0.5$ and $\alpha=\beta=0$).
\\
\bigskip
\\
The Gaussian distribution with mean $\mu$ and variance $\sigma^2$ are obtained for $\lambda=-0.5$, for $\alpha,\delta\rightarrow \infty$ and $\frac{\delta}{\alpha}\rightarrow \sigma^2$.\\ 
\bigskip
\\
The Skew-Student $t$ with $\nu$ degrees of freedom  is obtained if $\alpha=\vert \beta \vert$, then $\nu=-2\lambda>0$.\\
\bigskip
\\
The Student $t$ distribution is obtained for $\alpha=\beta=0$, $\mu=0$ and $\delta=\sqrt{\nu}$.
In the special case of hyperbolic distributions ($\lambda=1),$ we obtain the skewed Laplace distribution for $\delta=0.$
\\
\bigskip
\\
 All of these have been used to model financial returns and log-returns.
 \\
 \bigskip
 \\
  In \textsc{table} \ref{tab1}, we consider some specific \textit{Gh} distributions with the superscript $spe$ and limiting distributions with the superscript $lm$. 
 The used parameters are specified and the \textit{emef} figures corresponding to each choice are displayed. 

 \vspace{9ex}
\begin{center}
\begin{tabular}{|l| c c c c c|l|}
\hline 
\multirow{2}*{Distributions} & \multicolumn{5}{c|}{Parameters}& \\
\cline{2-6}
& $\lambda $  &  $\alpha $ & $\beta $& $\delta$& $\mu$ &\multirow{-2}*{Figures} \\
 \hline 
 Hyperbolique$^{spe}$ &$1$ &$1.5$  & $-0.5$ & $0.75$ & $0.2$ &  \\
 t-Stud.$^{spe}$  & $-2$ &  $10^{-8}$ & $0$  & $2$ & $0$ & \multirow{-2}*{Figure \ref{hypstud1}}\\
\multirow{2}*{Asym. t-Stud.$^{spe}$}  & $-1.278$ & $0.01186$ &$0.01186$  & $0.0766$ &1.005 &  \\
 & $-1.247$ & $0.0148$ &$-0.0147$  & $0.076$ & $1.005$ & \multirow{-2}*{Figure \ref{studstud1} }\\  
\multirow{2}*{NIG$^{spe}$}& -0.5 & 8.03 & -1.37&0.051 & 0.0105 &   \\ 
 & -0.5 & 7.6 &  -1.24& 0.052 & 0.0103 & \multirow{-2}*{ Figure \ref{nignig1} }\\ 
  Variance Gamma$^{spe}$&$2$&0.3 &0.1 &2 &0 &  \\
 $GIG^{spe}$ & $5 $& $3$  &  $1$ & - & - &  \multirow{-2}*{Figure  \ref{vggig1} }\\
  \hline  
\hline
  $IG$ Inverse Gaussian$^{lm}$ 
 &$0.5$ & $1$&$0$ & $1$ &$0$ &  \\
  Gamma$(\alpha,\beta)^{lm}$  &0.5& $4.5\, 10^{12}$ & $10^{-8}$
& -&- &\multirow{-2}*{Figure  \ref{invg1} } \\ 
$I\Gamma$ Inverse Gamma$^{lm}$   & $-0.5 $ & $1.9 \times 10^{-8}$ & $3.1 \times 10^{-3}$ & - & - &  \\ 
Skew Laplace$^{lm}$ &$1$ &$1.1$ &$0.1$ & $0.001$&$2$& \multirow{-2}*{Figure \ref{iglap1}}\\
 
Gaussian$^{lm}$($3,0.3)$ &$-0.5$&$10^{6}$ &2& $3\times 10^{5}$& 3& \\

Cauchy$^{lm}$ $(7,1)$ &$-0.5$&$0$  & $0$& 1&7 &\multirow{-2}*{Figure \ref{GausChy2} } \\
\hline
\end{tabular}
\vspace{3ex} 
\captionof{table}{Specific and limiting \textit{GH} distributions.}
\label{tab1}
\end{center} 

\begin{figure}
\includegraphics[scale=0.3]{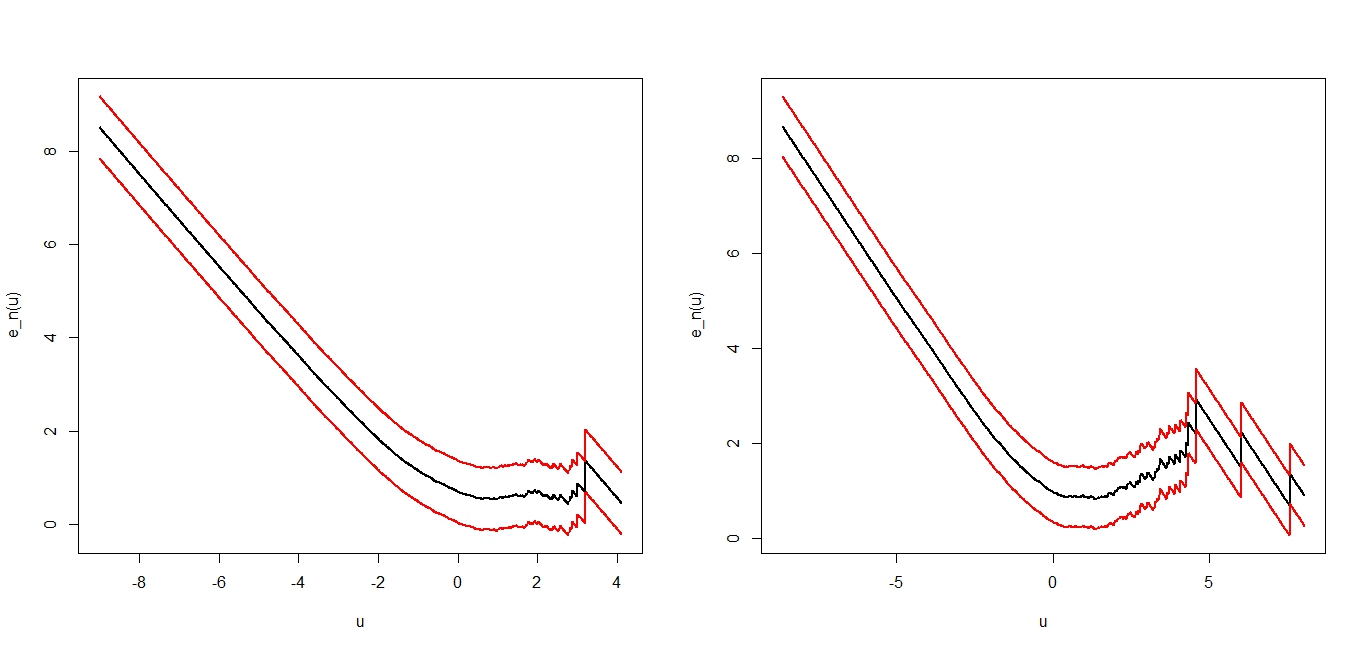}
\caption{ Left panel : \textit{Emef} for an hyperbolic distribution with the parameters $\lambda=1$, $\alpha=1.5$, $\beta=-0.5 $, $\delta=0.75$,  $\mu=0.2$. Right panel : \textit{Emef} for a $t$-student with the parameters $\lambda=-2$, $\alpha=10^{-8}$, $\beta=0 $, $\delta=2$,  $\mu=0$.}\label{hypstud1}
\end{figure}
\begin{figure}
\includegraphics[scale=0.3]{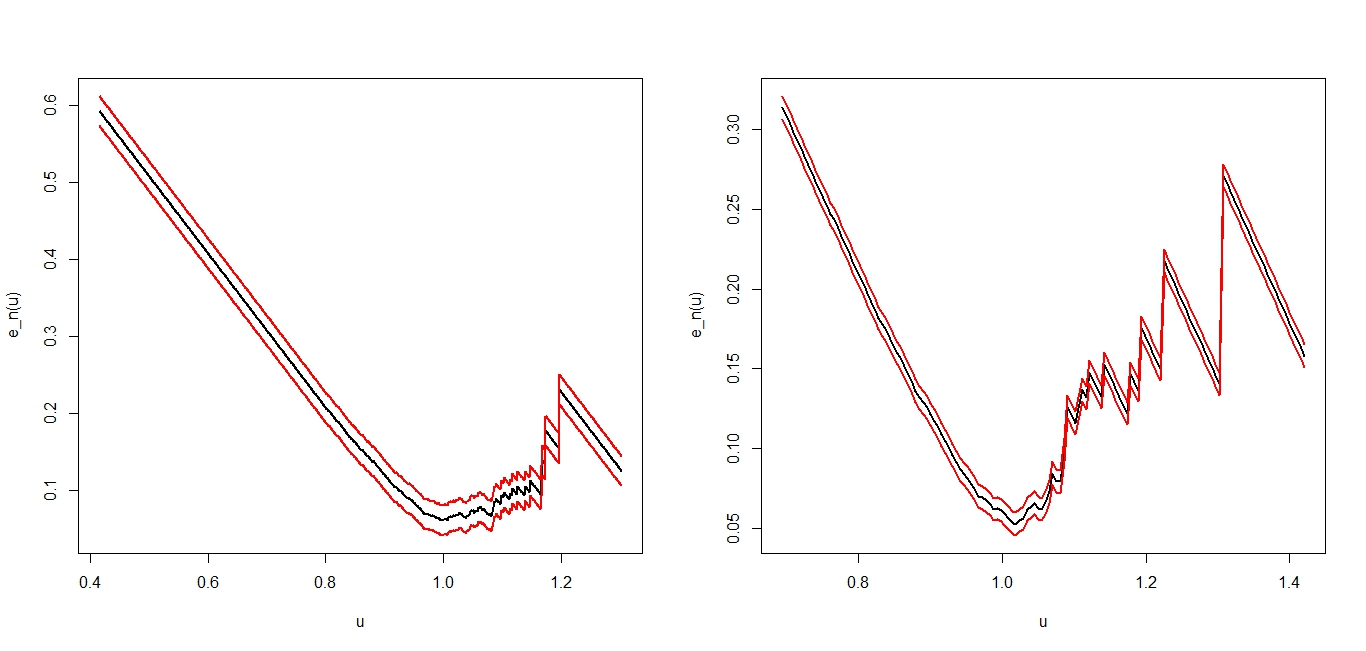} 
\caption{ The \textit{emef} for two $t$-Student distribution. The left panel concerns the one with the parameters $\lambda=-1.278$, $\alpha=0.01186$, $\beta=0.01186 $, $\delta=0.0766$,  $\mu=1.005$ and the right panel concerns the one with the parameters $\lambda=-1.247$, $\alpha=0.0148$, $\beta=-0.0148 $, $\delta=0.07683$,  $\mu=1.005$.}\label{studstud1}
\end{figure}
\begin{figure}
\includegraphics[scale=0.3]{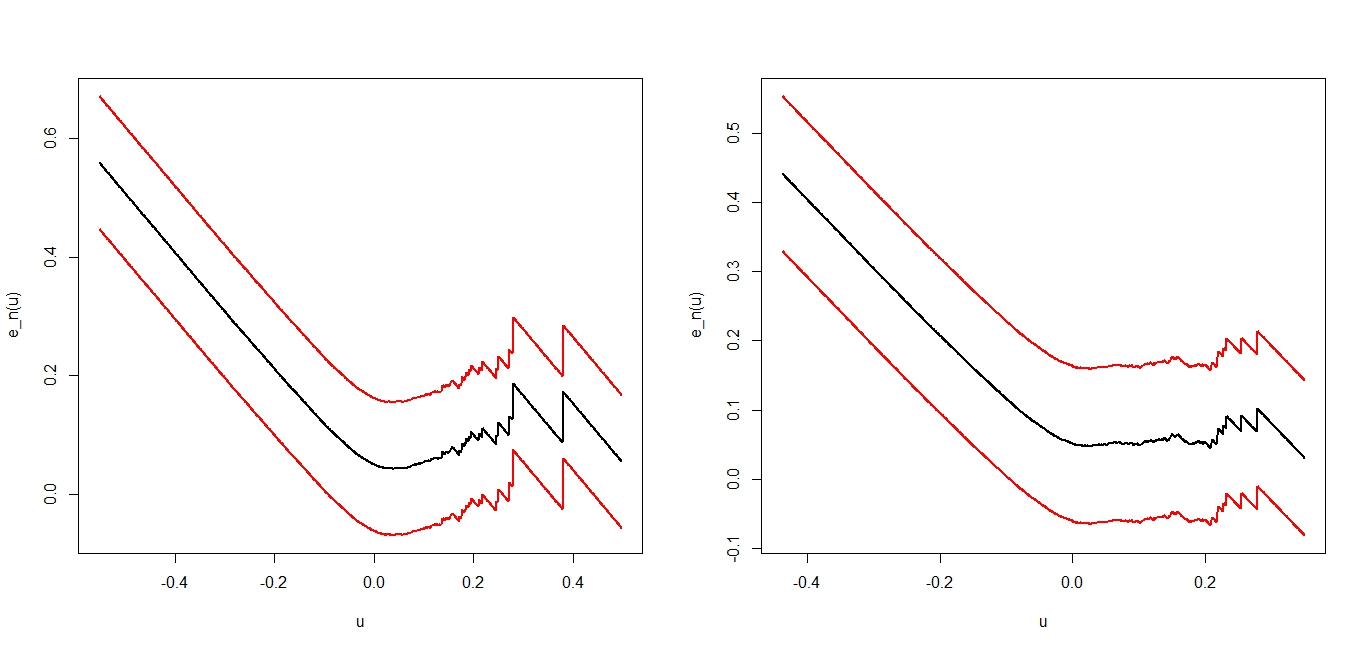} 
\caption{ The \textit{emef} for two Normal Inverse Gaussian distributions. The left panel concerns the one with the parameters $\lambda=-0.5$, $\alpha=8.03$, $\beta=-1.37 $, $\delta=0.051$,  $\mu=0.0105$ and the right panel concerns the one with the parameters $\lambda=-0.5$, $\alpha=7.6$, $\beta=-1.24 $, $\delta=0.052$,  $\mu=0.0103$.}\label{nignig1}
\end{figure}

\begin{figure}
\includegraphics[scale=0.3]{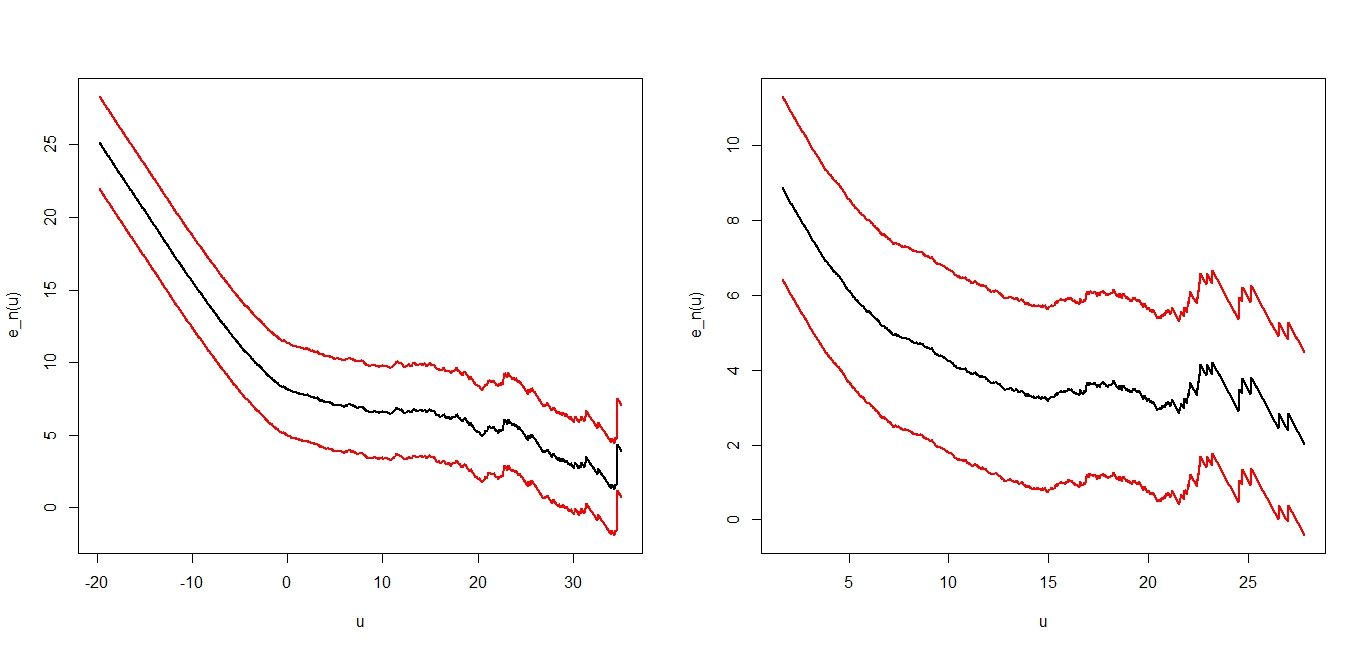} 
\caption{Left panel : the \textit{emef} for a Variance Gamma distribution with the parameters $\lambda=2$, $\alpha=0.3$, $\beta=0.1 $, $\delta=2$,  $\mu=0$. Right panel : the \textit{emef} for a Generalized Inverse Gaussian distribution with the parameters $\lambda=5$, $\alpha=3$, $\beta=1 $.}\label{vggig1}
\end{figure}

\begin{figure}
\includegraphics[scale=0.3]{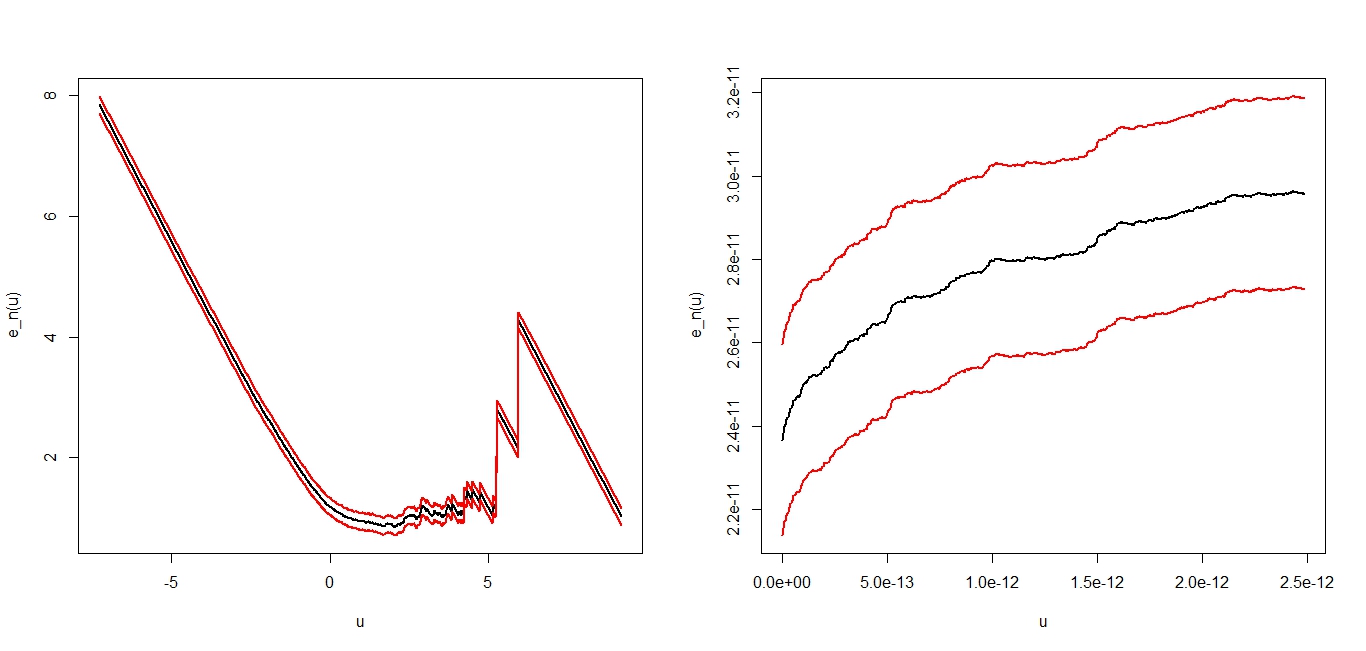} 
\caption{Left panel : the \textit{emef} for Inverse Gaussian distribution with the parameters $\lambda=0.5$, $\alpha=1$, $\beta=0$, $\delta=1$, $\mu=0$. Right panel : the \textit{Emef} for a Gamma distribution with the parameters $\lambda=0.5$, $\alpha=4.5\times 10^{12}$, $\beta=10^{-2} $.}\label{invg1}
\end{figure}
\begin{figure}
\includegraphics[scale=0.3]{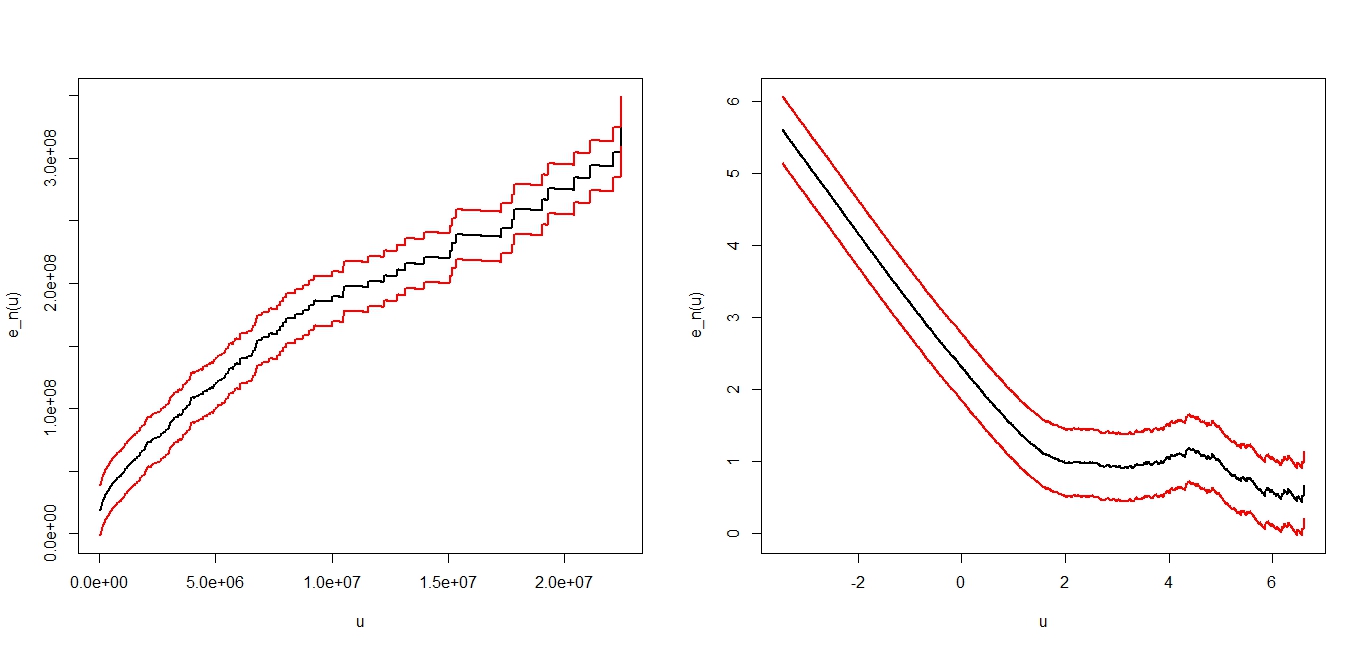} 
\caption{Left panel : the \textit{emef} for the Inverse Gamma distribution with the parameters \\ $\lambda=-0.5$, $\alpha=1.9\times 10^{-8}$, $\beta=3.1\times 10^{-3}. $ Right panel : the \textit{emef} for the Skew Laplace distribution with the parameters $\lambda=1$, $\alpha=1.1$, $\beta=0.1 $, $\delta=10^{-3}$,  $\mu=2$}\label{iglap1}
\end{figure}

\begin{figure}
 \includegraphics[scale=0.3]{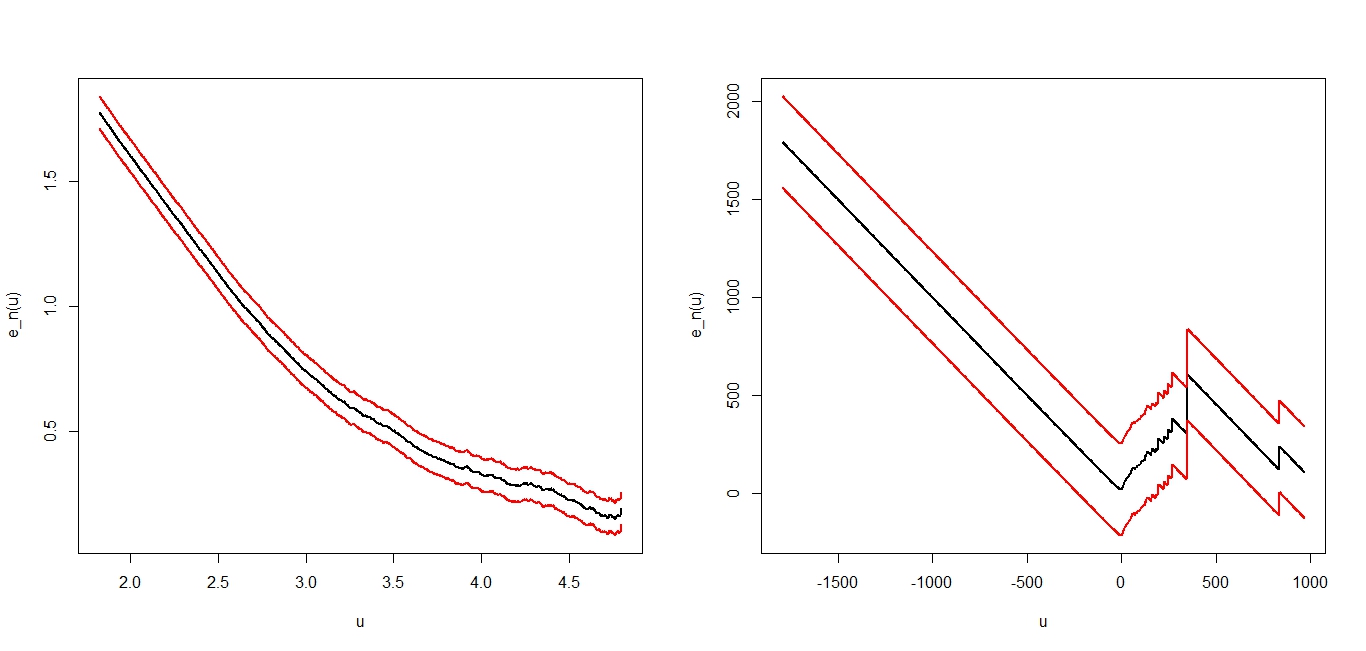} 
\caption{The left panel concerns the \textit{emef} for the Gaussian distribution with mean $\mu=2$ and variance $\sigma^2=0.3 $.
The right panel concerns the \textit{emef} for the Cauchy distribution with location parameter $\mu=7$ and scale parameter  $\delta=1$. }
\label{GausChy2}

\end{figure}
\subsubsection{Graphical test} We are now in a position to use the \textit{emef} graphs already drawn as tools of goodness of fit.
\\
\bigskip
\\
\textit{Emef} for 
Normal Inverse Gaussian (NIG) and $t$-student-distributions are not monotonic function. They decrease and increase like for \textit{emef} returns data. For this reason, 
we fit them to both monthly returns and log-returns from Dow Jones data base (see \textsc{figure} \ref{axprOMFit}, \textsc{figure} \ref{axprCMFit}, \textsc{figure} \ref{cscoreMVFitt}, and \textsc{figure} \ref{cscolrMVfit}). 
\\
\bigskip
\\
Dow Jones data base consists of several compagnies like AXP(American Express compagny), CSCO(Cisco Systems), DAX, CAT, IBM and so one. Each one having 5 values :  from opening (\textit{op}) values to closing (\textit{cl}) values , also minimum (\textit{min}), maximum (\textit{max}), and volume (\textit{vol}) values.\\
\bigskip
\\
We select AXP and CSCO compagnies and we consider returns and log-returns for their values as showed in the \textsc{table} \ref{tabb1}. Then we construct their \textit{emef} plot and their fitted counterpart.\\
\bigskip
\\
Estimates parameters and the \textit{emef} are given in \textsc{table} \ref{tabb2}.
\\
\bigskip

\vspace{4ex}\begin{center}
\begin{tabular}{|c|c|c|c|c|c|}
\hline 
 \multirow{2}*{Compagnies}  & \multirow{2}*{Nature} &\multirow{2}*{Values} &  \multicolumn{2}{c|}{\textit{Emef} plots} \\ 
\cline{4-5}
& & &\textit{Real emef}&\textit{Fitted emef}\\
\hline 
\multirow{2}*{AXP}
 & \multirow{2}*{Returns} &
  op  &  &\\
 \cline{3-3}
 & &min &\multirow{-2}*{Figure \ref{axprOM}}& \multirow{-2}*{Figure \ref{axprOMFit}} \\
 \cline{2-5}
 & \multirow{2}*{Log-returns.} &
  max  &  &\\
 \cline{3-3}
 & &cl &\multirow{-2}*{Figure \ref{axplrMC}}&\multirow{-2}*{Figure  \ref{axprCMFit}}\\
 \cline{2-5}
 \hline 
\multirow{2}*{CSCO}
 & \multirow{2}*{Returns} &
  min  &  &\\
 \cline{3-3}
 & &vol &\multirow{-2}*{Figure \ref{cscoreMV}}&\multirow{-2}*{Figure  \ref{cscoreMVFitt}}\\
 \cline{2-5}
 & \multirow{2}*{Log-returns} &
  op  &  &\\
 \cline{3-3}
 & &max &\multirow{-2}*{Figure \ref{cscologreMV}}&\multirow{-2}*{Figure  \ref{cscolrMVfit}}\\
 \cline{2-5}
 \hline 
\end{tabular}
\vspace{2ex}
\captionof{table}{(Fitted) \textit{Emef} for DAX and CSCO compagnies data.}
\label{tabb1}

\end{center}

\begin{figure}
\includegraphics[scale=0.3]{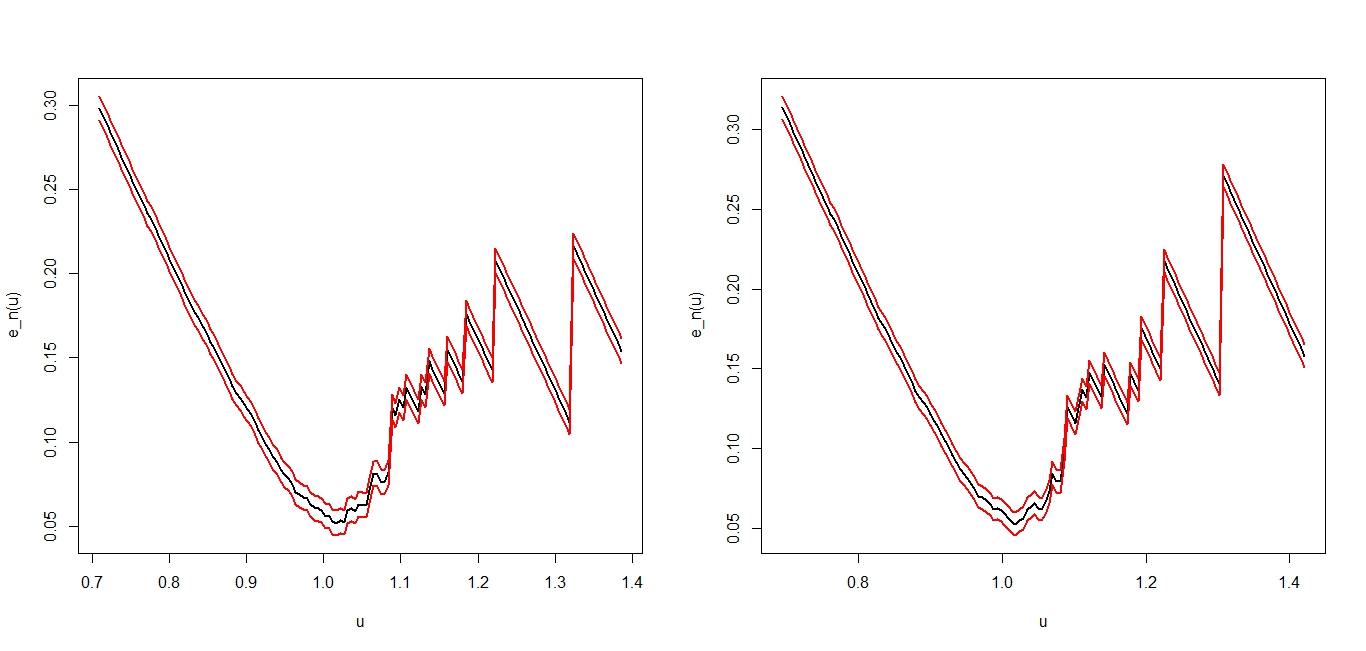} 
\caption{\textit{Emef} for AXP compagny (monthly data returns). The left panel concerns the opening values and the right one concerns the minimum values.}\label{axprOM}
\end{figure}

\begin{figure}
\includegraphics[scale=0.3]{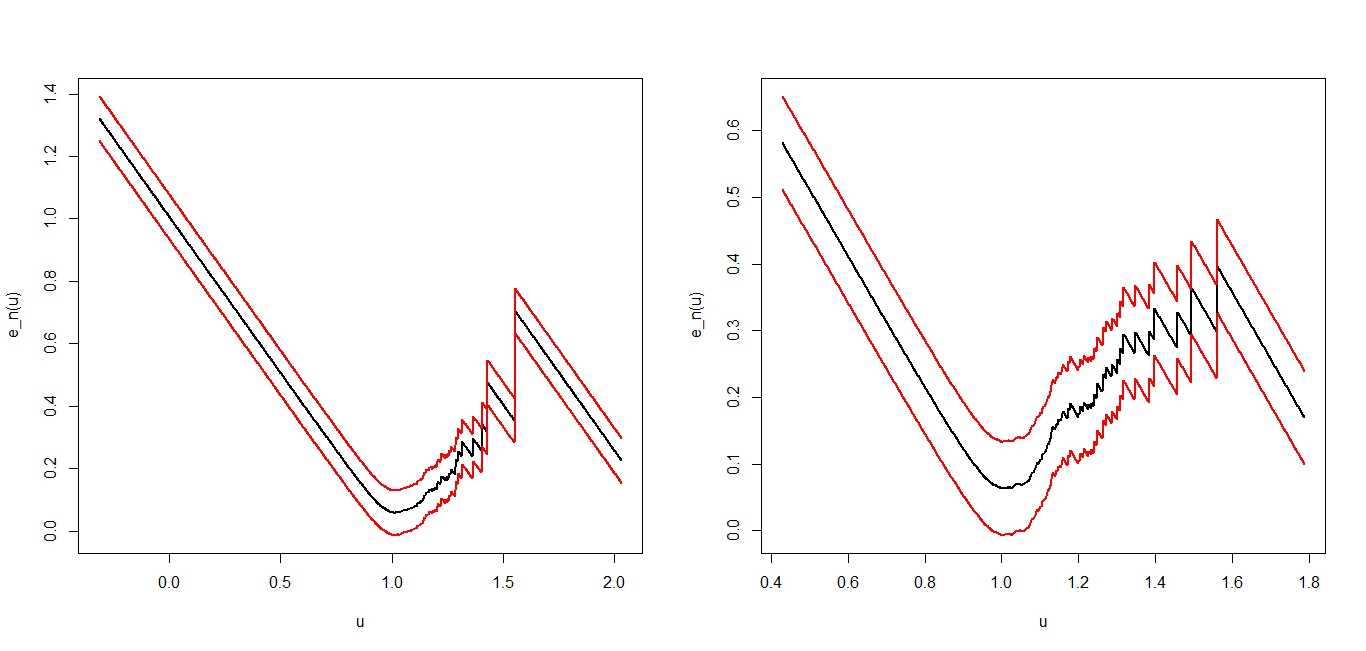} 
\caption{ A $t$-Student distribution is fitted to monthly data returns for AXP compagny (see figure \ref{axprOM}). The left panel concerns a $t$ distribution with the parameters $\lambda=-1.278,$\\ $\alpha=0.01186,\beta=0.01186,\delta=0.0766,$ and $\mu=1.005$ fitted to opening values.\\ The right one concerns 
a $t$ distribution with the parameters $\lambda=-1.247,\alpha=0.0148$,\\$\beta=-0.0148,\delta=0.07683,\mu=1.005$ fitted to minimum values.
 }\label{axprOMFit}
\end{figure}

\begin{figure}
\includegraphics[scale=0.3]{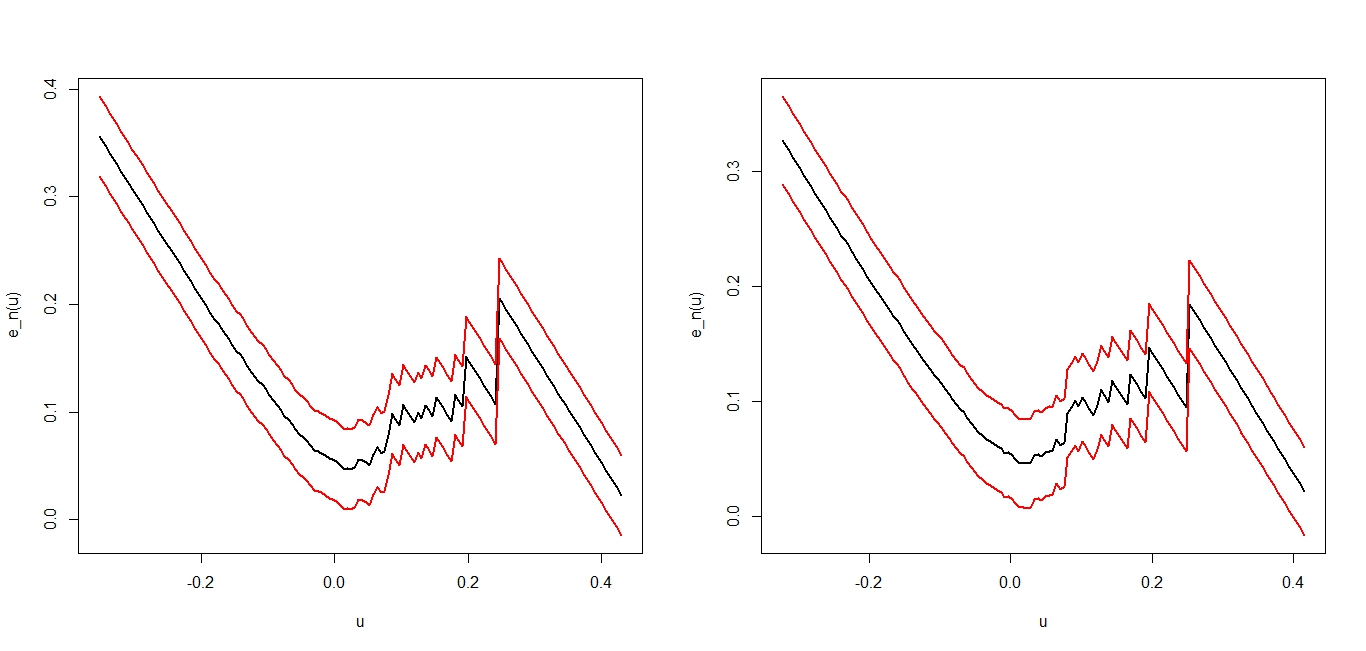} 
\caption{\textit{Emef} for AXP compagny (monthly data log-returns). The left panel concerns the maximum values and the right panel concerns the closing values.}\label{axplrMC}
\end{figure}

\begin{center}

\begin{tabular}{|c|c|c|c|c|c|c|c|c|l|}
\hline 
 \multirow{2}*{Comp}  & \multirow{2}*{Nature} &\multirow{2}*{Values} &  \multicolumn{5}{c|}{Ghyp estimates parameters }& &\\ 
\cline{4-8}
& & & $\hat{\lambda} $ & $\hat{\alpha }$ & $\hat{\beta} $& $\widehat{\delta}$ & $\hat{\mu}$  & \multirow{-2}*{ Fit.Dist }&\multirow{-2}*{ Figures }\\
\hline 
\multirow{2}*{AXP}					
 & \multirow{2}*{Returns} &\textit{op}.&$-1.278$
    &$0.01186$ & $0.0118$&$0.0766$ &$1.005$&$t$-stud &  \\
 \cline{3-9}
 & &\textit{min}. &-1.247 &0.0148 &-0.0148 &0.0768 &$1.005$ &$t$-stud &\multirow{-2}*{Fig. \ref{axprOMFit} }  \\
 \cline{2-10}
 & \multirow{2}*{Log-ret.} &
 \textit{ max}.  &-0.5  &8.03&-1.37 &0.051&0.0105 & NIG & \\
 \cline{3-9}
 & &\textit{cl}. & -0.5&7.6 &-1.24 &0.052 &0.0103& NIG &\multirow{-2}*{Fig. \ref{axprCMFit}}\\
 \cline{2-10}
 \hline 

\multirow{2}*{CSCO}
 & \multirow{2}*{Returns} &
  \textit{min}.  &-1.24 & 0.0148&-0.0148 &0.0768 &1 &$t$-Stud. & \\
 \cline{3-9}
 & &\textit{vol}. &-3.82&4.22&4.22 &0.613 &0.753 &$t$-Stud. & \multirow{-2}*{Fig. \ref{cscoreMVFitt}}\\
 \cline{2-10}
 & \multirow{2}*{Log-ret.} &
  \textit{op}. & -1.26 &0.83 &- 0.83&0.07 &0 &$t$-Stud.& \\
 \cline{3-9}
 & &\textit{max}. &-1.32 &0.85 &-0.85 &0.076 &0 &$t$-Stud. & \multirow{-2}*{Fig. \ref{cscolrMVfit}} \\
 \cline{2-10}
 \hline 
\end{tabular}
\vspace{2ex}
\captionof{table}{\textit{Emef} for fitted \textit{Gh} distributions to DAX and CSCO compagnies data.}
\label{tabb2}

\end{center}

\begin{figure}
\includegraphics[scale=0.3]{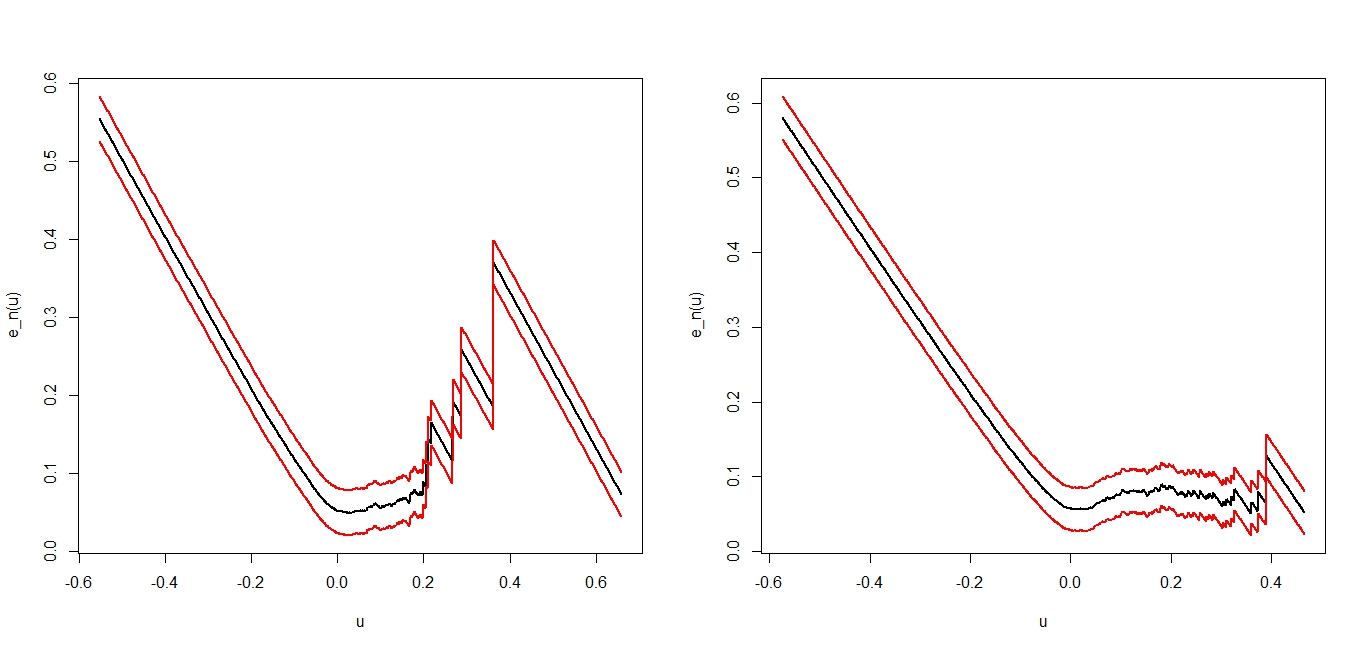} 
\caption{A NIG distribution is fitted to monthly data log-returns for AXP compagny (see Figure \ref{axplrMC}). The left panel concerns the one with the parameters  $\lambda=-0.5,\alpha=8.03,$\\ $\beta=-1.37,\delta=0.051,\mu=0.0105$ fitted to maximum values. The right panel concerns the one with the parameters $\lambda=-0.5,\alpha=7.6,\beta=-1.24,\delta=0.052,\mu=0.0103$ fitted to closing values.}\label{axprCMFit}
\end{figure}

\begin{figure}
\includegraphics[scale=0.3]{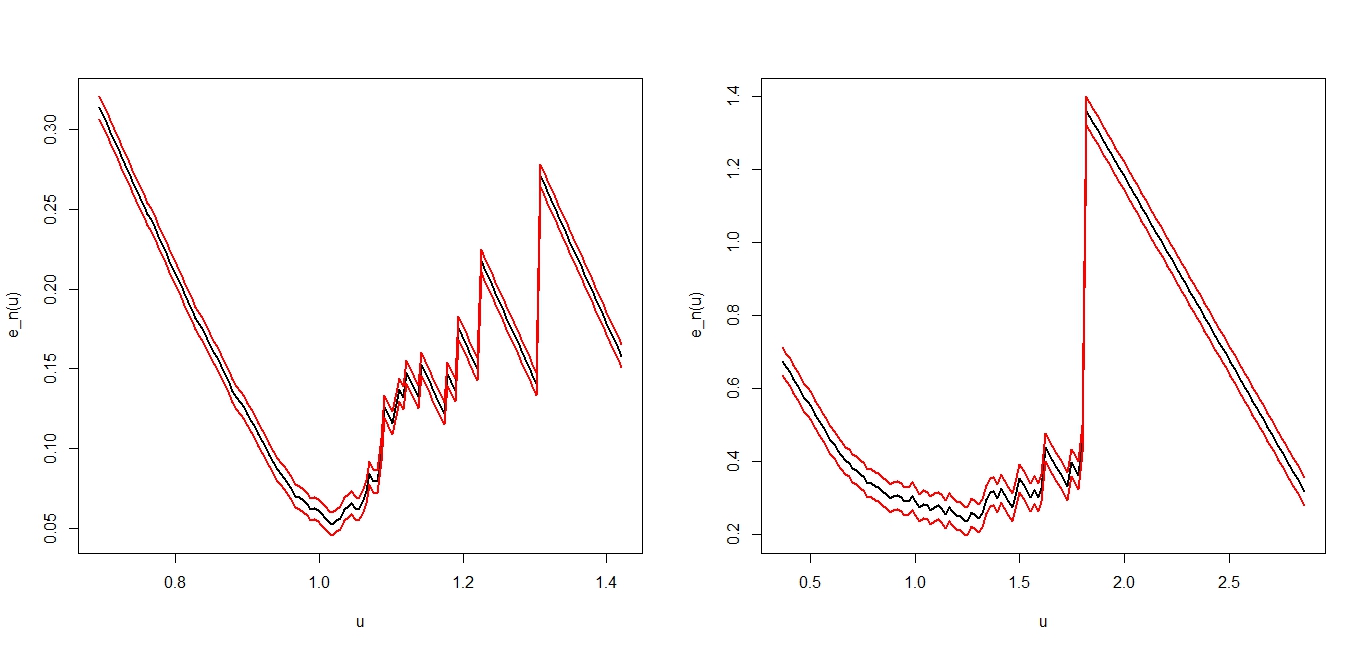}
 \caption{\textit{Emef} for CSCO compagny (data returns). The left panel concerns monthly minimum values and the right panel concerns  monthly volum values.}\label{cscoreMV}
\end{figure}

\begin{figure}

\includegraphics[scale=0.3]{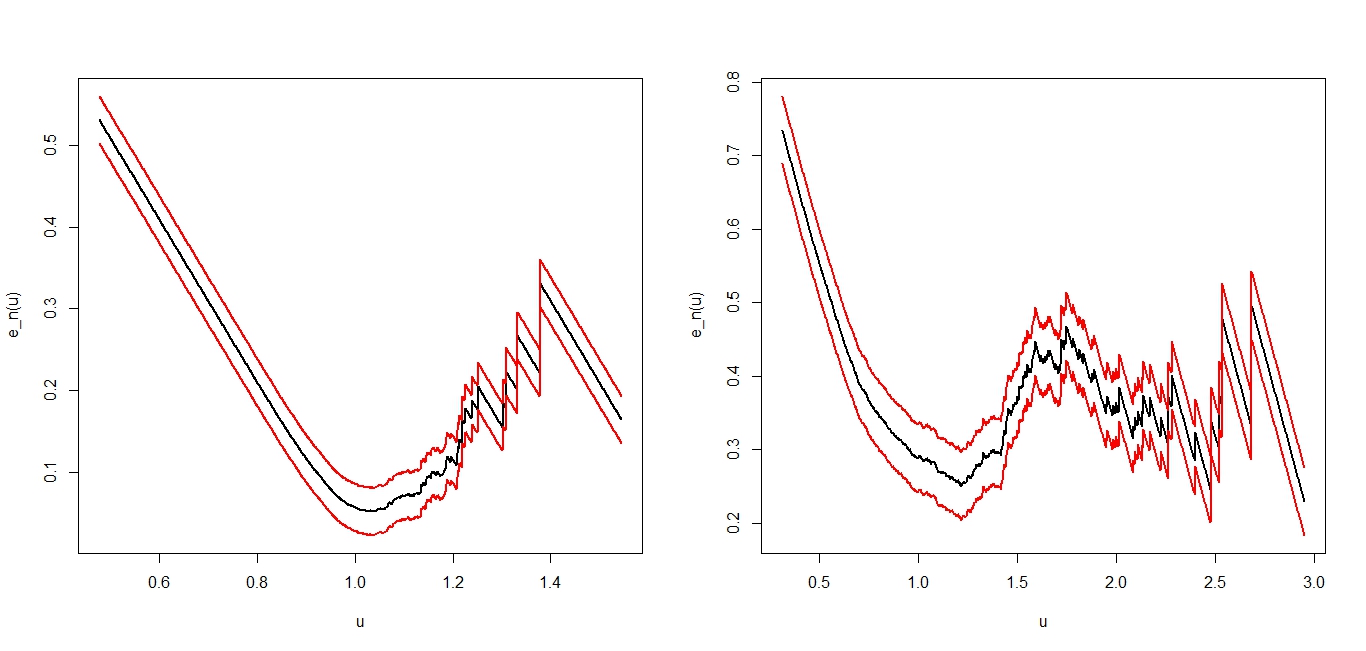} 
\caption{
A $t$-student distribution is fitted to monthly data returns for CSCO compagny (see Figure \ref{cscoreMV}). The left panel concerns the one with the parameters $\lambda=-1.24,\,\alpha=0.014,\\ \beta=-0.014,\,\delta=0.076,\,\mu=1$ fitted to minimum values. The right panel concerns the one with the parameters $\lambda=-3.82,\,\alpha=4.22,\beta=4.22,\,\delta=0.613,\,\mu=0.753$ fitted to volume values.}\label{cscoreMVFitt}
\end{figure}

\begin{figure}
\includegraphics[scale=0.3]{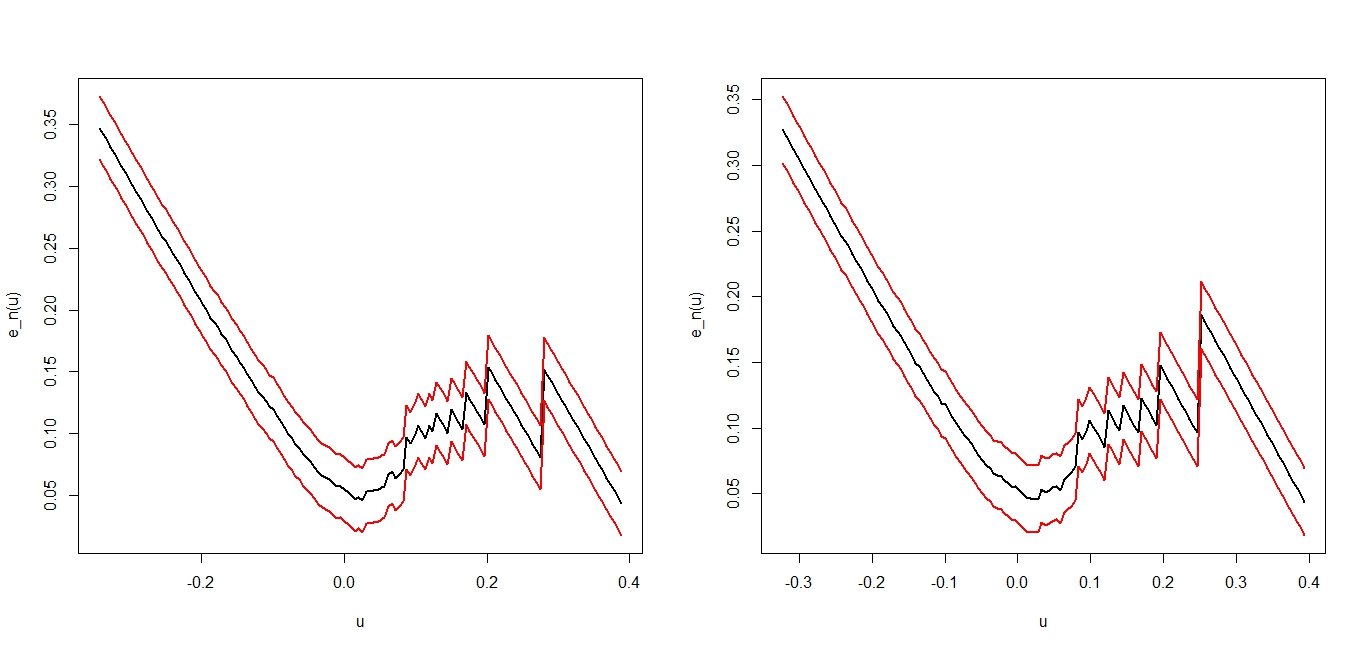} 
\caption{\textit{Emef} for CSCO compagny (data log-returns). The left panel concerns monthly opening values and the right panel concerns  monthly maximum values.}\label{cscologreMV}
\end{figure}  
  
  \begin{figure}
  
  \includegraphics[scale=0.3]{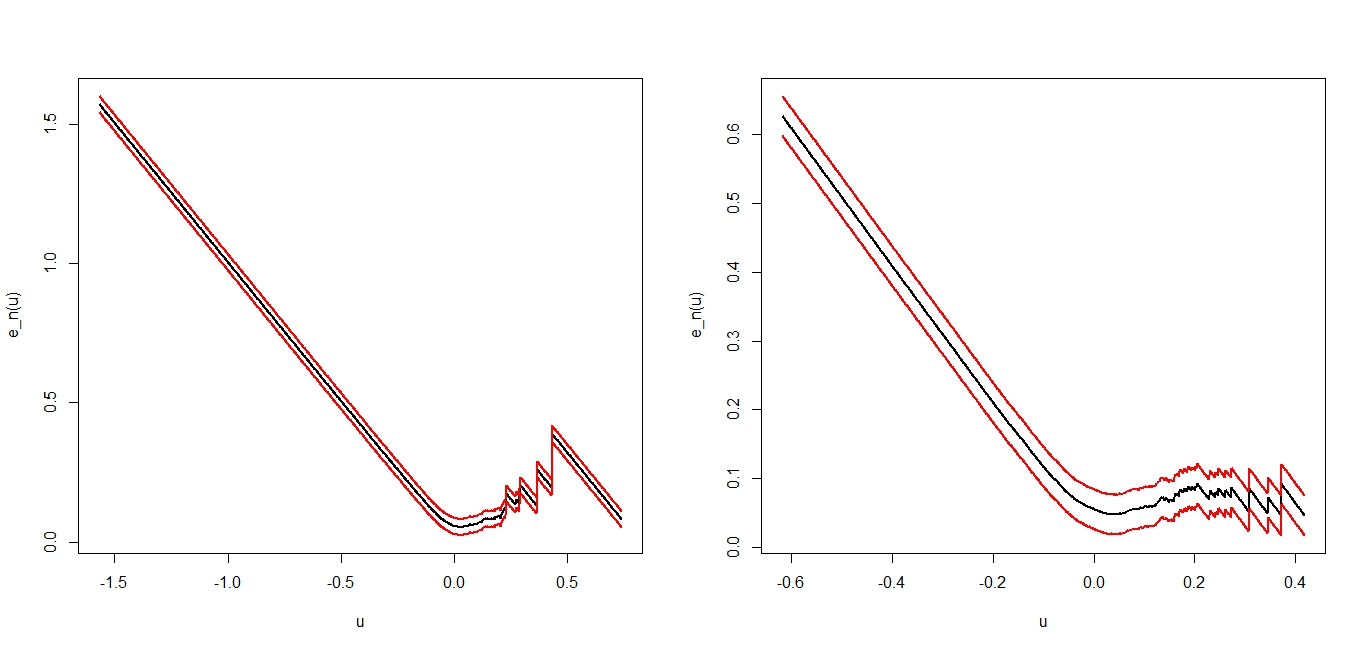} 
  \caption{A $t$-student distribution is fitted to monthly data log-returns for CSCO compagny (see Figure \ref{cscologreMV}). The left panel concerns the one with the parameters $\lambda=-1.26,\\ \alpha=0.83,\,\beta=-0.83,\,\delta=0.07,\,\mu=0$ fitted to opening values. The right panel concerns the one with the parameters $\lambda=-1.32,\,\alpha=0.85,\beta=-0.85,\,\delta=0.076,\,\mu=0$ fitted to maximum values.
  }\label{cscolrMVfit}
  \end{figure}
 
\subsubsection{Commentaries}  
In view of \textsc{figure} \ref{studstud1} and \textsc{figure} \ref{axprOMFit} we can say that $t$ studient distribution fits well opening and minimum values return for the American Express compagny AXP, whereas $NIG$ distribution fits well maximum and closing log-returns values for the Cysco System compagny CSCO in view of \textsc{figure} \ref{nignig1} and \textsc{figure} \ref{axprCMFit}. 
\newpage

\section{Conclusion \label{sec6}}
In this paper we have established an asymptotic confidence bands for the mean excess function by using functional process approach.
Then we applied these bands for fitting \textit{Gh} distributions to Dowjones financial data. It is a known fact that these ones fit well financial data since they embrace major part of classic distributions.

\bigskip

We remarked that Student and $NIG$ distributions are good candidates for fitting returns and log-returns data showing their semi-heavy tails. 
\\
\bigskip
\\
\section{Appendix}
\subsection{\protect\bigskip Moment computations}$\,$\\
\label{mocom}Let $Z_{1},...,Z_{n},$  $n$ i.i.d 
 centered random variables defined on the same
probability space with common variance $\mathbb{E}(Z_{i}^{2})=\kappa _{1}$ and common
fourth moment \ $\mathbb{E}(Z_{i}^{4})=\kappa _{2}>0$. We have

\begin{eqnarray}
\displaystyle \mathbb{E}\big[ T(n,u,\delta )\big] ^{4}&=&\mathbb{E}\Big(%
Z_1+Z_2+\ldots Z_n\Big) ^4  \notag \\
\notag &=&\mathbb{E}\Big(\sum_{k=1}^n Z_k^4+6 \sum_{1\leq i<j\leq n} Z_i^2 Z_j^2%
\Big) \\
&=&\sum_{k=1}^n \mathbb{E}(Z_k^4)+6 \sum_{1\leq i<j\leq n} \mathbb{E}(Z_i^2 Z_j^2)
 \label{Tn}
\end{eqnarray}%
since, for distinct $i$, $j$, $k$ and $l$, 
\begin{equation*}
\mathbb{E}(Z_i Z_j^3)=\mathbb{E}(Z_i Z_j^2 Z_k)=\mathbb{E}(Z_i Z_jZ_k Z_l)=0,
\end{equation*}
by using independence plus the fact that $\mathbb{E}(Z_i)=0$. 
 \\
Using independence again,  
\begin{equation*}
\mathbb{E}(Z_i^2 Z_j^2)=\mathbb{E}(Z_i^2)\mathbb{E}( Z_j^2)=\kappa _{1}^2 \ \ \mbox{for} \ \ i\neq j.
\end{equation*}We obtain $\displaystyle \sum_{1\leq i<j\leq n} \mathbb{E}(Z_i^2 Z_j^2)=\frac{n(n-1)}{2} \kappa_1^2$  since the number of possible couples $(i,j)$ of integers such that $1\leq i<j\leq n$, is $\displaystyle \binom{n}{2}= \frac{n(n-1)}{2}$.
Hence from \eqref{Tn}, we deduce that 
\begin{equation*}
\displaystyle \mathbb{E}\big[ T(n,u,\delta )\big] ^{4}=n \kappa
_{2}+3n(n-1)\kappa _{1}^2.
\end{equation*}

\begin{flushright}
$\square$
\end{flushright}

\subsection{\protect\bigskip Proofs of the uniform asymptotic consistency
bounds.}
\subsubsection{Talagrand bounds}
\label{unifasc}
We begin to recall the Talagrand bounds and a device of Einmahl and Mason on
how to apply it.\\
 Before going any further, we recall that a class of measurable
real valued functions $\mathcal{F}$ is said to be a \textit{pointwise
measurable class} if there exists a countable subclass $\mathcal{F}_{0}\ $ of 
$\mathcal{F}$ such as, for any function $f$ in $\mathcal{F}$, we can find a
sequence of functions $\{f_{m}\}_{m\geq 0}$ in $\mathcal{F}_{0}$ for which $%
f_{m}(x)\rightarrow f(x) \ \ \mbox{as} \ \ m \rightarrow \infty$ , $x\in \mathbb{R}$. (See Example 2.3.4 in \cite{vaart}.
 \\ Further, let $\xi _{1},\xi _{2},\ldots $ be a sequence of
independent Rademacher random variables independent of $X_{1},X_{2},\ldots,$ and $\mathbb{G}_m$ be the functional empirical process indexed by the class of functions $\mathcal{F}$. \\
\bigskip
\\
The following inequality is essentially due to Talagrand (1994) (see \cite%
{talagrand}).
\\
\bigskip
\\
\textbf{Inequality.} Let $\mathcal{F}$ be a \textit{pointwise measurable class%
} of functions satisfying for some \\$0<M < \infty, \, \Vert f\Vert
_{\infty }\leq M,\, \ \ f\in \mathcal{F},$ .\\ Then for all $t>0$ we have, 
\begin{eqnarray}\label{ineq}
&&\mathbb{P}\Big\{\max_{1\leq m\leq n}\Vert \sqrt{m}\mathbb{G}_{m}\Vert _{%
\mathcal{F}}\geq A_{1}\Big(\mathbb{E}\Big\|\sum_{i=1}^{n}\xi _{i}f(X_{i})%
\Big\|_{\mathcal{F}}+t\Big)\Big\}  \notag \\
\ \ \ \ \  && \ \ \ \ \ \ \ \ \ \ \ \ \ \  \leq   \ \  2\big(\exp (-A_{2}\,t^{2}/n\sigma _{\mathcal{F}}^{2})+\exp
(-A_{2}\,t/M)\big),
\end{eqnarray}%
where $\sigma _{\mathcal{F}}^{2}=\sup_{f\in \mathcal{F}}Var(f(X))$ and  $A_{1},A_{2}$ are universal constants.\\
\bigskip\\
 And the
lemma below of Einmahl and Mason \cite{EM2010} is very helpful for obtaining
bounds on this quantity, when the class $\mathcal{F}$ has a polynomial
covering number. \par
Assume that there exists a finite valued measurable
function $G$, called an envelope function, which satisfies for all $\displaystyle x\in 
\mathbb{R},\,G(x)\geq \sup_{f\in \mathcal{F}}\left\vert f(x)\right\vert $.
We define for $0<\epsilon <1$ 
\begin{equation*}
N(\varepsilon ,\mathcal{F}):=\sup_{Q}N\Big(\epsilon \sqrt{Q(G^{2})},\mathcal{%
F},d_{Q}\Big)
\end{equation*}%
where the supremum is taken over all probability measures $Q$ on $\mathbb{R}$
for which  $0<Q(G^{2}):=\int G^{2}(y)Q(dy)<\infty $ and $d_{Q}$ is the $%
L_{2}(Q)-$metric. As usual, $N(\epsilon ,\mathcal{F},d_{Q}$) is the minimal
number of balls $\{g:d_{Q}(g,f)<\epsilon \}$ of $d_{Q}-$radius $\epsilon $
needed to cover $\mathcal{F}$. Here is the device of Einmahl and Mason \cite%
{EM2010}.

\begin{lemma}
\label{lemEM}(Einmahl - Mason \cite{EM2010}) Let $\mathcal{F}$ be a
pointwise measurable class of bounded functions such that for some constants 
$\beta>0 ,\,\nu>0 ,\,C>1,\,\sigma \leq 1/(8C)$ and function $G$ as above, the
following four conditions hold:

\begin{enumerate}
\item[(A.1)] $\mathbb{E}\Big[G^{2}(X)\Big]\leq \beta ^{2};$

\item[(A.2)] $N(\epsilon ,\mathcal{F})\leq C\varepsilon ^{-\nu },\ \
0<\epsilon <1$;

\item[(A.3)] $\displaystyle \sigma _{0}^{2}:=\sup_{f\in \mathcal{F}}\mathbb{E}\Big[f^{2}(X)%
\Big]\leq \sigma ^{2};$

\item[(A.4)] $\displaystyle \sup_{f\in \mathcal{F}}\Vert f\Vert _{\infty }\leq \frac{1}{2%
\sqrt{\nu +1}}\sqrt{n\sigma ^{2}/\log (\beta \vee 1/\sigma )}.$%

\end{enumerate}
Then we have for some absolute constant $A$,
\begin{equation}
\mathbb{E}\Big\|\sum_{i=1}^{n}\xi _{i}f(X_{i})\Big\|_{\mathcal{F}}\leq A%
\sqrt{\nu n\sigma ^{2}\log (\beta \vee 1/\sigma )}.  \label{em2001}
\end{equation}
\end{lemma}

\subsubsection{\textbf{APPLICATION}}
Put $\ell_u(x)=\ell(x)\mathbb{I}_{(x > u)}$, with $\ell(x)=1$ or $\ell(x)=x$,
 and let $\mathcal{F}=\{\ell _{u},u\in I\}.$  \newline
$\mathcal{F}$ is pointwise measurable since it suffices to take $\mathcal{F}%
_{0}=\{\ell_u, u\in I\cap \mathbb{Q}\},$ where $\mathbb{Q}$ is the set of
irrationnal numbers.
\\
 \newline
Next $G=\max (\left\vert \ell (u_{0})\right\vert ,\left\vert \ell(u_{1})\right\vert )=M>0$ is an envelope of $\mathcal{F}$ since we have  $$\displaystyle \sup_{u \in I}|\ell_u(x)|\leq |\ell(x)|\leq \max (\left\vert \ell (u_{0})\right\vert ,\left\vert \ell(u_{1})\right\vert),\quad \forall \, u_0\leq x\leq u_1.$$ 
Remark that if $\ell(x)=1$ then $M=1$ and if $\ell(x)=x$ then $M=\max(|u_0|,|u_1|).$\\
We have $\displaystyle \sigma _{\mathcal{F}}^{2}= \sup_{f\in \mathcal{F}%
}Var(f(X)) \leq M^2.$\newline
So we may use Talagrand's inequality : it remains to check points of \textbf{Lemma \ref{lemEM}}:\\
 Points (A.1) and (A.3) are obvious with $\beta=M =\sigma.$
\\ To check (A.2), consider any probability $Q$ on $\mathbb{R}.$
We get for $(u,v)\in I^{2},u\leq v,$%
\begin{equation}
d_{Q}^{2}(\ell _{u},\ell _{v})=\int (\ell _{u}-\ell _{v})^{2}(x)dQ(x)\leq
M^{2}Q([u,v]).  \label{ent}
\end{equation}%
By a classical result in probability in $\mathbb{R}$, 
for any given $0<\varepsilon <1,$ we may cover $[u_{0},u_{1}]$ by at most \\ $\displaystyle m=\lceil \frac{u_1-u_0}{\varepsilon} \rceil$ sub-intervals $\displaystyle [s_{i-1},s_i]$ such that 
$Q([s_{i-1},s_i])<\varepsilon^2, \ \ i=1,\ldots,m.$\\
($\lceil x\rceil$ stands for the smallest positive integer greater than or equal to $x$). \\ Let $C=(m+1)\varepsilon$, we have $m<C\varepsilon^{-1}$.\\ 
\bigskip
\\
For any $u \in [u_0, u_1]$, there exists $i\in \{1,\ldots,m\}$ such as $s_{i-1}\leq u \leq s_i$ with $Q([u,s_i])<\varepsilon^2$, so the corresponding 
$\ell_u \in \mathcal{F}$ is such that  $$d_Q(\ell_u,\ell_{s_{i}})<\varepsilon M=\varepsilon \sqrt{Q(G^2)} \ \ \ \mbox{from \eqref{ent}}.$$
To finish $m=N\Big(\epsilon \sqrt{Q(G^{2})},\mathcal{F},d_{Q}\Big)<C\varepsilon^{-1}$ and 
\begin{equation*}
N(\epsilon ,\mathcal{F})=\sup_Q N\Big(\epsilon \sqrt{Q(G^{2})},\mathcal{%
F},d_{Q}\Big) \leq C\varepsilon ^{-1}.
\end{equation*}%
\\
\bigskip
\\
Now we take $\beta ^{2}=\sigma ^{2}=\max(2,\max (\left\vert \ell
(u_{0})\right\vert ,\left\vert \ell (u_{1})\right\vert )=M_{1}.$\\
 Finally for 
\begin{equation*}\displaystyle n\geq \frac{8 M^2\log M_1}{M_1^2}
,\end{equation*}%
we have
\begin{equation}
\mathbb{E}\Big\|\sum_{i=1}^{n}\xi _{i}g(X_{i})\Big\|_{\mathcal{F}}\leq C_{\mathcal{F}}%
\sqrt{n},  \label{app}
\end{equation}%
where
 $\displaystyle C_{\mathcal{F}}=A\,M_1\sqrt{\log M_1},$ since all the points of the \textbf{Lemma \ref{lemEM}} are checked. \\
\bigskip
\\Now we are going to apply the inequality \eqref{ineq} first for the class of functions  $$\mathcal{F}_{1}=\{\ell _{u}(x)=g_{u}(x),u\in I \}.$$ 
In this case $M_1=2$ since $\ell(x)=1$, for any $u_0\leq x \leq u_1$, and $$\mathbb{E}\Big\|\sum_{i=1}^{n}\xi _{i}g_u(X_{i})\Big\|_{\mathcal{F}_1}= D_{n,1}\leq  C_{\mathcal{F}_1}\sqrt{n}, \ \ \mbox{where} \ \ C_{\mathcal{F}_1}=2A \sqrt{\log 2}. $$
Let $\varepsilon>0,\,n_1 \geq 2 \log 2$  and $t_0$ such that  $$\exp \Big(\frac{-A_{2}t_{0}^{2}}{n_1} \Big)\leq \frac{\varepsilon }{8},\ \ \mbox{and}\ \ \exp \Big(-A_{2}t_{0} \Big)\leq \frac{\varepsilon }{8}\ \ \mbox{and}\ \ \ t_0<\sqrt{n_1}.$$(Remind that $\sigma_{\mathcal{F}}^2=1$.)\\ Then
\begin{eqnarray*}
\displaystyle \mathbb{P}\Big\{\max_{1\leq m\leq n} \| \sqrt{m}\mathbb{G}_m\|_{%
\mathcal{F}_1}\geq A_{1}\Big(\mathbb{E}\Big\|\sum_{i=1}^{n}\xi _{i}g_u(X_{i})%
\Big\|_{\mathcal{F}_1}+\, t_0\Big)\Big\} \leq \varepsilon /2.
\end{eqnarray*}%

So for $n\geq n_{1},$ we arrive at%
\begin{equation*}
\mathbb{P}\Big( \left\vert \mathbb{P}_{n}(g_{u})-\mathbb{P}_{X}(g_{u})\right\vert < \frac{A_{1}(D_{n,1}+t_{0})}{n},u\in I 
\Big)> 1-\varepsilon /2.
\end{equation*}%
As $t_0/\sqrt{n}<\sqrt{n_1}/\sqrt{n}\leq1$, we obtain $$\frac{A_{1}(D_{n,1}+t_{0})}{n}\leq \frac{A_{1}C_{\mathcal{F}_1} \sqrt{n}+A_1 \sqrt{n}}{n}=\frac{A_1C_{\mathcal{F}_1}+A_1}{\sqrt{n}}= \frac{D_1}{\sqrt{n}},$$ thus
\begin{equation}  \label{cbgu}
\mathbb{P}\Big( \left\vert \mathbb{P}_{n}(g_{u})-\mathbb{P}_{X}(g_{u})\right\vert < \frac{D_1}{\sqrt{n}} ,u\in I 
\Big)> 1-\varepsilon /2
\end{equation}%
where $D_{1}= 2A\,A_1 \sqrt{\log 2}+A_1$.
\\
\bigskip
\\

Let us use the same method, for the class of functions $$ \mathcal{F}_{2}=\{ \ell _{u}(x)=f_{u}(x),u\in I \}.$$
In this case $M_1=\max(2,\max(|u_0|,|u_1|))$ since $\ell(x)=x,$ for any $u_0\leq x \leq u_1$, and 
 $$
\mathbb{E}\Big\|\sum_{i=1}^{n}\xi _{i}f_u(X_{i})\Big\|_{\mathcal{F}_2}=D_{n,2} \leq C_{\mathcal{F}_2}\sqrt{n}, \mbox{where} \ \ C_{\mathcal{F}_2}=AM_1\sqrt{\log M_1}.$$
Let $\displaystyle n_2 \geq \frac{8 M^2\log M_{1}}{M_{1}^2}$  and $t_0$ such that  $$\exp \Big(\frac{-A_{2}t_{0}^{2}}{n_2} \Big)\leq \frac{\varepsilon }{8}\ \ \mbox{and}\ \ \exp \Big(-A_{2}t_{0} \Big)\leq \frac{\varepsilon }{8}\ \ \mbox{and}\ \ t_0<\sqrt{n_2}.$$ Then
\begin{eqnarray*}
\displaystyle \mathbb{P}\Big\{\max_{1\leq m\leq n} \| \sqrt{m}\mathbb{G}_m\|_{%
\mathcal{F}_2}\geq A_{1}\Big(\mathbb{E}\Big\|\sum_{i=1}^{n}\xi _{i}f_u(X_{i})%
\Big\|_{\mathcal{F}_2}+t_0\Big)\Big\} \leq \varepsilon /2.
\end{eqnarray*}%

So for $n\geq n_{2},$ we deduce that%
\begin{equation*}
\mathbb{P}\Big( \left\vert \mathbb{P}_{n}(f_{u})-\mathbb{P}_{X}(f_{u})\right\vert < \frac{A_{1}(D_{n,2}+t_{0})}{n},u\in I 
\Big)> 1-\varepsilon /2.
\end{equation*}%
As $t_0/\sqrt{n}<\sqrt{n_1}/\sqrt{n}\leq1$, we obtain $$\frac{A_{1}(D_{n,2}+t_{0})}{n}\leq \frac{A_{1}C_{\mathcal{F}_2} \sqrt{n}+A_1 \sqrt{n}}{n}=\frac{A_1C_{\mathcal{F}_2}+A_1}{\sqrt{n}}= \frac{D_2}{\sqrt{n}},$$ thus
\begin{equation}  \label{cbfu}
\mathbb{P}\Big( \left\vert \mathbb{P}_{n}(f_{u})-\mathbb{P}_{X}(f_{u})\right\vert < \frac{D_2}{\sqrt{n}} ,u\in I 
\Big)> 1-\varepsilon /2,
\end{equation}%
where $D_{2}= A\,A_1\,M_{1} \sqrt{\log M_{1}}+A_1$.
\\
\bigskip 
\\ 
 Now we use again (\ref{ingen}) : 
$$|{e}_{n}(u)-e(u)| \leq  |\mathbb{P}_{n}(f_{u})-\mathbb{P}_{X}(f_{u})|\times |\mathbb{P}_{n}(g_{u})|^{-1}  \ + \ |\mathbb{P}%
_{X}(f_{u})| \times \frac{|\mathbb{P}_{n}(g_{u})-\mathbb{P}_{X}(g_{u})|}{|%
\mathbb{P}_{n}(g_{u})\mathbb{P}_{X}(g_{u})|}. $$
For $u_0\leq u \leq u_1$,  
 we get \begin{equation*}
0<\overline{F}(u_1)-\frac{D_1}{\sqrt{n}}\leq \mathbb{P}_{X}(g_{u})-\frac{D_1}{\sqrt{n}}<
\mathbb{P}_{n}(g_{u})<\mathbb{P}_{X}(g_{u})+\frac{D_1}{\sqrt{n}}
\end{equation*} with a probability greater than (\textit{w.p.g.t}) $1-\varepsilon/2$ 
and thus 
\begin{equation*}
\displaystyle |\mathbb{P}_{n}(g_{u})|^{-1}<  \Big(\overline{F}(u_1)-\frac{D_1}{\sqrt{n}}\Big)^{-1}\ \ \mbox{\textit{w.p.g.t}}\ \ 1-\varepsilon/2
\end{equation*}
so we obtain
 \begin{eqnarray*}   
&&\mathbb{P}\left( |\mathbb{P}_{n}(f_{u})-\mathbb{P}_{X}(f_{u})|\times |\mathbb{P}_{n}(g_{u})|^{-1} < \frac{D_2}{\sqrt{n}} \times \Big( \overline{F}(u_1)-\frac{D_1}{\sqrt{n}}\Big)^{-1} ,u\in I \right)  \\
&&   \ \ \  \ \ \ \ \ \  \ \ \ \  \ \ \ \geq \mathbb{P}\Big( \left\vert \mathbb{P}_{n}(f_{u})-\mathbb{P}_{X}(f_{u})\right\vert <  \frac{D_2}{\sqrt{n}} ,u\in I \Big) > 1-\varepsilon/2,
\end{eqnarray*}
thus \begin{equation}\label{ef1}\mathbb{P}\left( |\mathbb{P}_{n}(f_{u})-\mathbb{P}_{X}(f_{u})|\times |\mathbb{P}_{n}(g_{u})|^{-1} < \frac{D_2}{\sqrt{n}} \times \Big( \overline{F}(u_1)-\frac{D_1}{\sqrt{n}}\Big)^{-1} ,u\in I \right)> 1-\varepsilon/2.
 \end{equation}From the following inequalities :  $$ \begin{cases}\displaystyle |\mathbb{P}_{n}(g_{u})|^{-1}
\leq \Big(\overline{F}(u_1)- \frac{D_1}{\sqrt{n}} \Big)^{-1},\ \ \mbox{\textit{w.p.g.t}}\ \ 1-\varepsilon/2,\\
|\mathbb{P}_{X}(g_{u})|^{-1}\leq \overline{F}(u_1)^{-1},\\
 |\mathbb{P}_{X}(f_{u})|\leq \mathbb{E}|X|,
\end{cases}
 $$and by the same manner, we obtain \begin{eqnarray*}
&&\mathbb{P}\left( |\mathbb{P}_{X}(f_{u})| \times \frac{|\mathbb{P}_{n}(g_{u})-\mathbb{P}_{X}(g_{u})|}{|%
\mathbb{P}_{n}(g_{u})\mathbb{P}_{X}(g_{u})|}<\frac{D_1}{\sqrt{n}}\times \mathbb{E}|X|\times \Big( \overline{F}(u_1)(\overline{F}(u_1)-\frac{D_1}{\sqrt{n}}) \Big)^{-1}, u \in I\right)\\
&&   \ \ \  \ \ \ \ \ \  \ \ \ \  \ \ \ \geq \mathbb{P}\Big( \left\vert \mathbb{P}_{n}(g_{u})-\mathbb{P}_{X}(g_{u})\right\vert <  \frac{D_1}{\sqrt{n}} ,u\in I \Big) > 1-\varepsilon/2 \,,
 \end{eqnarray*}
thus \begin{equation}\label{ef2}
 \mathbb{P}\left( |\mathbb{P}_{X}(f_{u})| \times \frac{|\mathbb{P}_{n}(g_{u})-\mathbb{P}_{X}(g_{u})|}{|%
\mathbb{P}_{n}(g_{u})\mathbb{P}_{X}(g_{u})|}<\frac{D_1}{\sqrt{n}}\times \mathbb{E}|X|\times \Big( \overline{F}(u_1)(\overline{F}(u_1)-\frac{D_1}{\sqrt{n}}) \Big)^{-1}\right)> 1-\varepsilon/2
\end{equation}
 By combining \eqref{ef1} and \eqref{ef2}, we obtain \begin{eqnarray*}
&& \mathbb{P}\left( |\mathbb{P}_{n}(g_{u})|^{-1}\times\left\vert \mathbb{P}_{n}(f_{u})-\mathbb{P}_{X}(f_{u})\right\vert \geq \frac{D_2}{\sqrt{n}} \times\Big( \overline{F}(u_1)-\frac{D_1}{\sqrt{n}}\Big)^{-1} ,u\in I \right)\\ 
&& \ \ \ \  \ \ \  \ \ \ \ \ \   \ \ \ \ +   \ \ \ \mathbb{P}\left( |\mathbb{P}_{X}(f_{u})| \times \frac{|\mathbb{P}_{n}(g_{u})-\mathbb{P}_{X}(g_{u})|}{|%
\mathbb{P}_{n}(g_{u})\mathbb{P}_{X}(g_{u})|} \geq \frac{D_1}{\sqrt{n}}\times \mathbb{E}|X|\times \Big( \overline{F}(u_1)(\overline{F}(u_1)-\frac{D_1}{\sqrt{n}}) \Big)^{-1}\right)\\
&& \ \ \ \  \ \ \  \ \ \ \ \ \   \ \ \ \ \leq  \ \ \ \frac{\varepsilon}{2} \ \ \ \  + \ \ \ \ \frac{\varepsilon }{2} \ \ \ \ \leq \ \ \ \ \varepsilon\,.
 \end{eqnarray*}
This gives 
 $$\mathbb{P}(|e_n(u)-e(u)|\geq \frac{E_n}{\sqrt{n}})\leq \varepsilon \ \ \ \mbox{where} \ \ \ E_{n}=\frac{1}{\overline{F}(u_{1})-D_{1}/\sqrt{n}}\Big(D_{2}+ \frac{D_{1}\times\mathbb{E}\vert X \vert}{ 
\overline{F}(u_{1})}\Big).$$ Finally we conclude by :
\begin{equation*}
\mathbb{P}\Big(e_{n}(u)-\frac{E_{n}}{\sqrt{n}}<e(u)<e_n(u)+\frac{E_{n}}{%
\sqrt{n}},u\in I\Big)> 1-\varepsilon
\end{equation*}for any $\varepsilon>0$, any $n\geq n_0
$  with  
\begin{equation*}
E_{n}=\frac{1}{\overline{F}(u_{1})-D_{1}/\sqrt{n}}\Big(D_{2}+ \frac{D_{1}\times\mathbb{E}\vert X \vert}{\overline{F}(u_{1})}\Big) \ \ \ \mbox{and} \ \ \ \ \ \left\{ 
\begin{array}{l}
\displaystyle D_{1}= 2A A_1 \sqrt{\log 2}+A_1 \cr \displaystyle D_{2}= A\,A_1\,M_{1} \sqrt{\log M_{1}}+A_1 ,\cr%
M_1=\max (2 ,\max(\vert u_0 \vert,\vert u_1 \vert))\cr
\end{array}%
\right.
\end{equation*} 
 \begin{flushright}
$\square$
\end{flushright}

\end{document}